%% file: main.tex
\documentclass[screen,sigplan]{acmart}
\renewcommand\footnotetextcopyrightpermission[1]{} 
\AtBeginDocument{%
  }

\usepackage{amsmath,amsfonts}
\usepackage{algorithmic}
\usepackage{algorithm}
\usepackage{graphicx}
\usepackage{textcomp}
\usepackage{xcolor}
\usepackage{hyperref}
\usepackage{subcaption}
\usepackage{multirow}
\usepackage{comment}
\usepackage{listings}

\definecolor{codegreen}{rgb}{0,0.6,0}
\definecolor{codegray}{rgb}{0.5,0.5,0.5}
\definecolor{codepurple}{rgb}{0.58,0,0.82}
\definecolor{backcolour}{rgb}{0.95,0.95,0.92}

\lstdefinestyle{mystyle}{
    backgroundcolor=\color{backcolour},   
    commentstyle=\color{codegreen},
    keywordstyle=\color{magenta},
    numberstyle=\tiny\color{codegray},
    stringstyle=\color{codepurple},
    basicstyle=\ttfamily\footnotesize,
    breakatwhitespace=false,         
    breaklines=true,                 
    captionpos=b,                    
    keepspaces=true,                 
    numbers=left,                    
    numbersep=5pt,                  
    showspaces=false,                
    showstringspaces=false,
    showtabs=false,                  
    tabsize=2
}
\begin{document}

\title{MPC-Pipe: An Efficient Pipeline Scheme for Semi-honest MPC Machine Learning}

\author{Yongqin Wang}
\email{yongqin@usc.edu}
\affiliation{%
 \institution{University of Southern California}
 \city{Los Angeles}
 \state{CA}
 \country{USA}}

\author{Rachit Rajat}
\email{rrajt@usc.edu}
\affiliation{%
 \institution{University of Southern California}
 \city{Los Angeles}
 \state{CA}
 \country{USA}}

\author{Murali Annavaram}
\email{annavara@usc.edu}
\affiliation{%
 \institution{University of Southern California}
 \city{Los Angeles}
 \state{CA}
 \country{USA}}

\newcommand{\fixme}[1]{
  \noindent 
  \colorbox{yellow} {\scriptsize FIXME}
  {\bf[}\textcolor{red}{#1}{\bf]} 
}
\newcommand{\revision}[1]{#1}
\newcommand{\sheperd}[1]{#1}
\newcommand{\pcent}[1]{#1\%}

\begin{abstract}
\input{0.abstract}
\end{abstract}

\maketitle
\pagestyle{plain}

\section{Introduction}
\label{sec:introduction}
\input{1.introduction}

\section{Related Works}
\label{sec:related}

\input{2.related-works}

\section{Background}
\label{sec:background}
\input{3.background}

\section{MPC-Pipe}
\label{sec:pipe}
\input{4.mpcpipe}

\section{Evaluation}
\label{sec:evaluation}
\input{5.evaluation}

\section{Conclusion}
\label{sec:conclusion}
\input{6.conclusion}

\section*{Acknowledgment}
We sincerely thank all the reviewers for their time and constructive comments. This material is based upon work supported by Defense Advanced Research Projects Agency (DARPA) under Contract Nos. HR001120C0088, NSF award number  2224319, REAL@USC-Meta center, and VMware gift. The views, opinions, and/or findings expressed are those of the author(s) and should not be interpreted as representing the official views or policies of the Department of Defense or the U.S. Government.

\bibliographystyle{plain}
\bibliography{references}

\end{document}

%% file: 0.abstract.tex
Multi-party computing (MPC) has been gaining popularity as a secure computing model over the past few years. However, prior works have demonstrated that MPC protocols still pay substantial performance penalties compared to plaintext, particularly when applied to ML algorithms. The overhead is due to added computation and communication costs. Prior studies, as well as our own analysis, found that most MPC protocols today sequentially perform communication and computation.
The participating parties must compute on their shares first and then perform data communication to allow the distribution of new secret shares before proceeding to the next computation step. In this work, we show that serialization is unnecessary, particularly in the context of ML computations (both in Convolutional neural networks and in Transformer-based models). We demonstrate that it is possible to carefully orchestrate the computation and communication steps to overlap. 

We propose MPC-Pipe, an efficient MPC system for both training and inference of ML workloads, which pipelines computations and communications in an MPC protocol during the online phase. MPC-Pipe proposes three pipeline schemes to optimize the online phase of ML in the semi-honest majority adversary setting. The three pipeline schemes are 1) inter-linear pipeline,  2) inner-layer pipeline, and 3) inter-batch pipeline. Inter-linear pipeline focuses on linear layers; inner-layer pipeline focuses on non-linear layers; inter-batch pipeline focuses on communication and computation overlaps in different input batches. We implement MPC-Pipe by augmenting a modified version of CrypTen, which separates online and offline phases. We evaluate the end-to-end system performance benefits of the online phase of MPC using deep neural networks (VGG16, ResNet50) and Transformers using different network settings. We show that MPC-Pipe can improve the throughput and latency of ML workloads.

%% file: 1.introduction.tex
Machine Learning (ML) is gaining popularity in many fields like health care, finance, and advertisement because ML models can glean knowledge from a large amount of data. Recently, ML providers have relied on cloud-based servers to perform ML model inference and training. Inputs that the user provides to ML models can be privacy-sensitive features, such as their voice signatures or their personal images. Similarly, model parameters may also be proprietary and need to be protected. But in a cloud environment, both the model and user data are vulnerable to a wide attack surface consisting of compromised hypervisors, physical snooping, and more~\cite{snoop1,hashemi2022data}. Secure Multi-Party Computation (MPC) is a potential solution to this challenge. MPC allows users and model owners to share their data and model parameters securely with mutually distrustful parties and train or serve a model. This process can ensure the confidentiality of both weights and input data.

\noindent \textbf{Our MPC setting:} In this work, we study the n-party semi-honest majority setting, which ensures the confidentiality of data against a semi-honest adversary that can corrupt up to $t<n$ of the parties. A semi-honest adversary will not introduce errors to the MPC computation but can still be curious about the data provided by MPC clients. The ``majority'' in the ``semi-honest majority'' means that the semi-honest adversary can corrupt $N/2$ or more MPC servers. For the protocol this paper focuses on, the semi-honest adversary can corrupt at most $N-1 \geq N/2$ parties without compromising data confidentiality.

There are two common MPC protocols: 1) Garble Circuit-based (GC-based) MPC~\cite{yao}, and 2) secret sharing-based (SS-based) MPC~\cite{secretshareop}. In this paper, we focus on the MPC protocols using secret shares.  GC-based MPC has fewer rounds of communication, and hence, they benefit in a setting where the network latency is large. SS-based MPC protocols have the advantage of less communicated bytes and less computation costs. Moreover, SS-based MPC protocols can scale more efficiently to more parties. 

\noindent \textbf{Online-offline phases in MPC:} \sheperd{MPC protocol adopts an online/offline computation paradigm. The online phase tasks can only be performed once the inputs are available. Offline phase tasks are input-independent and can be performed ahead of time.  This separation is advantageous because offline phase tasks, which tend to be computationally intensive tasks, can be completed beforehand, thereby removing the offline tasks from the critical path. Faster online phases are particularly desirable because they improve the protocol's responsiveness, making MPC more suitable for practical deployment. As such, this work focuses on improving the online phase.}

\subsection{Key observations about MPC frameworks}

\textbf{Communication \& computation serialization:} While MPC provides privacy, MPC protocols suffer significant performance costs compared to the plaintext~\cite{wangCharacIspass}. MPC protocols require MPC servers to perform additional computation and communication. In most MPC protocols, the participating parties first operate on their local shares of operands and then broadcast data that will not leak the original input to other servers (more details in the next section). This communication step is necessary to create new operands in the next computation phase. Thus, MPC protocols go through a series of compute-communication phases.

\begin{figure}[h]
  \centering
  \begin{subfigure}[tb]{0.35\linewidth}
    \includegraphics[width=\linewidth]{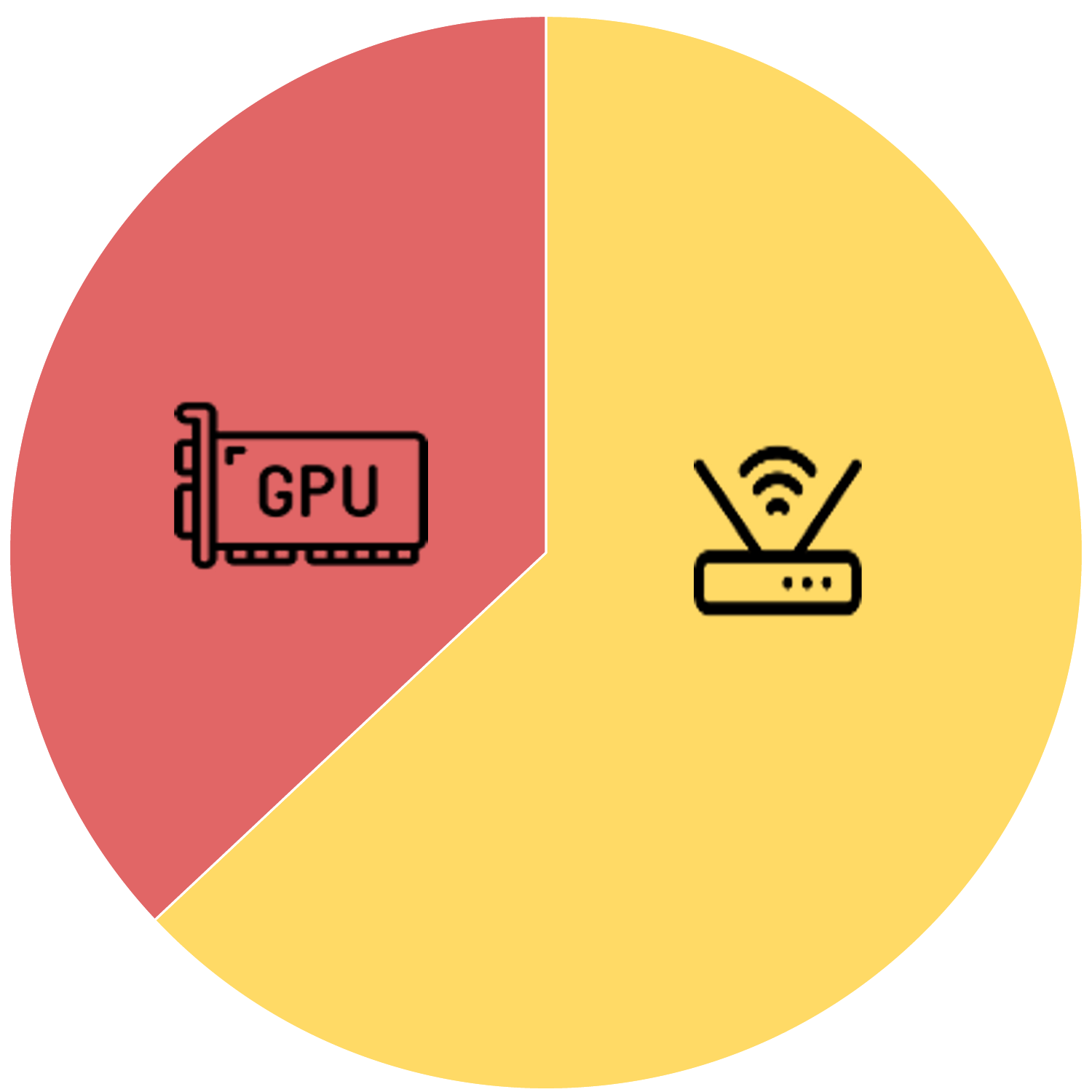}
     \caption{ResNet}
  \label{fig:resnet-movtivation}
  \end{subfigure}
  \begin{subfigure}[tb]{0.35\linewidth}
    \includegraphics[width=\linewidth]{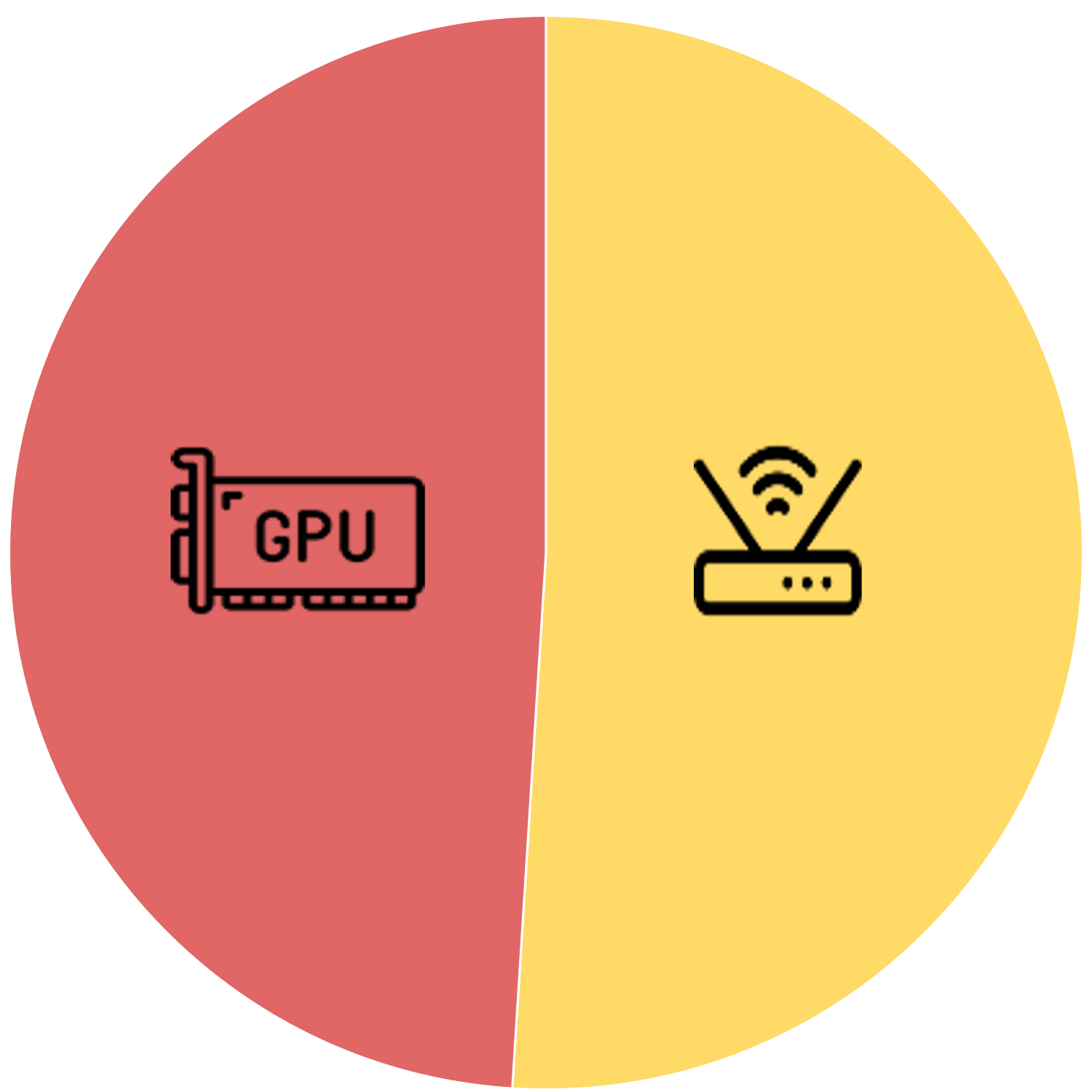}
    \caption{Transformer}
  \label{fig:xformer-motivation}
  \end{subfigure}
  \caption{MPC ML model communication and computation decomposition; the GPU icon represents computation runtime, and the router icon represents communication runtime.}
\label{fig:motivation}
\end{figure}

We make the observation that in current popular MPC frameworks~\cite{kumar2020cryptflow, delphi, mp-spdz, gazzele, chameleon, secureml, fis}, the computation and communication phases are serialized, and both communication and computation APIs are blocking calls. Intuitively, such serialization seems necessary so that MPC parties first wait for the new operands to be communicated before they can perform computations on those new operands. However, this means that an MPC server's compute fabric (usually GPUs) is underutilized while awaiting communication. 
\\


\noindent \textbf{The cost of serialization :} \revision{MPC protocols, when applied to ML workloads, suffer significant serialization delays between computation and communication.  We measured the compute and communication time breakdowns for a 2PC (with 2 MPC servers) inference setting with VGG16 and Transformer. Figure~\ref{fig:motivation} shows the fraction of time the GPUs and network are used. Due to serialization, the network channels are idle when GPUs are computing, and during the communication phase, the GPUs are idle.
The GPUs on MPC servers are idle for $63\%$ and $56\%$ of the total inference time on Resnet and Transformer models, respectively. Most of the GPU idleness occurs during the Softmax or ReLU operations, which have been demonstrated to be communication intensive~\cite{wangCharacIspass}. Similarly, when MPC servers perform computations, communication channels between MPC servers are idle, especially for computation-intensive ML layers such as Convolution and Dense layers. These layers account for about 45\% of the total runtime, and during this time window, communication channels are idle.} 

\noindent \textbf{Computation bottlenecks in ML MPC:} Our measurements on ML workloads demonstrate that computation costs are still substantial. Further analysis showed three reasons for the significant compute time spent by ML workloads in MPC: 1) MPC increases total computations by tripling the matrix multiplication when using beaver triples (see Section~\ref{sec:beaverTriple}), 2) MPC operands must be represented as fixed-point, and due to insufficient long fixed-point kernel support for GPUs, such fixed-point operations are slow to compute.  ~\sheperd{In fact, state-of-the-art GPU MPC implementations~\cite{crypten2020, fis} break long fixed-point numbers into smaller blocks and encode them into multiple floating-point operations to utilize the highly optimized floating point CUDA kernels. Even with such an optimized implementation, the fixed-point operations in linear operations become compute-dominant rather than communication-dominant}. In particular, a single fixed point multiplication takes 10 total floating-point multiplications due to this breaking up of a fixed point into smaller operations. 3) The computation complexity of MPC protocols for matrix multiplication and convolution is higher than communication complexity. Suppose MPC parties were to compute a matrix multiplication using matrix multiplication triples between two matrices of size $M\times N$ and $N\times O$. In that case, the computation complexity is $M\times N \times O$ while the communication complexity is $M\times N + N\times O$ (Section~\ref{sec:beaverTriple}). Thus, linear layers in ML, like convolution and matrix multiplication, have significantly more computation than communication. All those three reasons combined make computation hurdles equally noticeable for ML workloads when using MPC protocols.

Given that compute is a significant fraction of the total runtime, we make a case for breaking the serialization between computation and communication in MPC protocols when applied to ML workloads. In particular, we show that some data dependencies in current MPC protocol implementations are unnecessary. For example, before MPC parties execute an ML model's forward or backward pass, many operands needed during forward and backward pass are available much earlier than needed for computation. During the forward pass, model weights of all layers are available even before the computation propagates across layers. During the backward pass, intermediate activation (computed during the forward pass) and model weights are also available. We exploit the early availability of those operands to break communication and computation serialization and remove false dependencies by pipelining computation and communication in MPC. Our pipelined approach is termed as MPC-Pipe, and it improves resource utilization and reduces latency.  

Although the concept of overlapping computation and communication is well-established, it is important to study this overlap in the context of MPC, a crucial privacy-preserving ML framework. As ML grapples with privacy concerns, optimizing MPC (along with other privacy-preserving frameworks) is essential.  Using an in-depth understanding of ML workloads and MPC, this work identifies and leverages opportunities for domain-specific acceleration of privacy-preserving MPC. This approach echoes similar advancements within the distributed training community, as evidenced by several contributions in the field~\cite{dist1, dist2, dist3}.

MPC protects model parameters and inputs without making any assumptions regarding the underlying hardware capabilities, such as trusted execution enclaves~\cite{tramer2019slalom, darkNight}. Thus, MPC is a very desirable solution for deployment when data protection is a key requirement.  

\subsection{Our Contribution}
MPC-Pipe makes the following major contributions:
\begin{itemize}
  \item We identify the false data dependencies between communication and computation in secret-shared-based MPC when executing  ML layers, including Dense, Convolution, ReLU, MaxPooling, and Softmax.
  \item \sheperd{In general, the computation and communication runtime is skewed differently in linear and non-linear ML layers in MPC. Hence, the extent of this skew will limit overlapping computation and communication.} We propose three pipeline schemes to maximize the opportunities to overlap computation and communication in MPC ML.:
  \begin{enumerate}
      \item \textbf{inter-linear pipeline } overlaps computation and communication while performing linear algebraic computations on model weights, such as Convolutions and Dense. \sheperd{Linear layers are generally computation bound, and hence, the overlap is constrained by the amount of communication delay observed in the linear layers.} 
      \item \textbf{inner-layer pipeline} allows overlapping computation and communication in non-linear ML layers. \sheperd{Non-linear layers are generally communication bound, and hence, the overlap is constrained by the amount of computation delay observed in the non-linear layers.} 
      \item \textbf{inter-batch pipeline} allows pipelining across 2 different input batches. \sheperd{This scheme allows us to overlap the computations in linear layers of one batch with the communication needs of the non-linear layers in a different input batch.} 
  \end{enumerate} 
  \item We evaluate MPC-Pipe using CNN (VGG16~\cite{vgg} and ResNet50~\cite{resnet}) and NLP models (Transformers used in XLM-R). We also show MPC-Pipe's performance benefit in both local area networks (LAN) and wide area networks (WAN) with various interconnection speeds. In our experimental setting, MPC-Pipe can improve MPC model workload throughput by up to 50$\%$ and latency by up to 16$\%$. Note that all the throughput and latency gains in this work are achieved without requiring any MPC protocol changes or additional hardware modifications. Hence, the results presented are full end-to-end system improvements that can be achieved in the current generation hardware, which is an additional appealing aspect of our work. \sheperd{Note that because MPC-Pipe overlaps computation and communication to gain performance improvement, the effectiveness of each pipeline scheme will be determined by how communication and computation are distributed in each layer/model. Thus, different pipeline schemes will contribute differently to overall performance improvement, as we will see in later sections.}     
  \item We also analyze MPC-Pipe's performance benefits as we scale to more parties.
\end{itemize}


%% file: 2.related-works.tex
Several SS-based MPC frameworks have been proposed in the literature for 2PC and 3PC-specific settings~\cite{aby,kumar2020cryptflow, delphi, mp-spdz, gazzele, chameleon, secureml, fis}. Frameworks like Delphi~\cite{delphi},  Gazelle~\cite{gazzele}, XONN~\cite{xonn}, and SecureML~\cite{secureml} are specific for 2PC only settings.  CrypTFlow~\cite{kumar2020cryptflow}, ABY3~\cite{aby3}, and CryptGPU~\cite{fis} are specifically for 3PC settings. The SS protocol we are using is built on top of CrypTen~\cite{crypten2020} and can be applied to an arbitrary number ($N$) of parties tolerating $N-1$ corrupted parties. Falcon~\cite{wagh2020falcon} and CrypTFlow~\cite{kumar2020cryptflow} demonstrate privacy-preserving ML on ImageNet. However, they do not fully incorporate the MPC computation stage into GPUs. Further, these prior works do not address the compute-communication serialization issues, which is the primary focus of this work. There are works that address challenges in MPC protocols with malicious adversaries~\cite{spdz, mascot, overdrive, spdz2k, spdk2kml, compacttag}. Our work focuses on semi-honest adversaries.


{There are prior works that try to reduce the communication cost of MPC protocols~\cite{kss, covertMPC}. However, their method merges individual communication between \textit{CPU threads} to reduce the number of communication rounds. Our vectorized GPU implementation of MPC ML operations inherently achieves such merging by large vector broadcast and parallel computation.}

\subsection{MPC Operation Optimizations}
Sphynx \cite{reluMPC1}, DeepReduce \cite{reluMPC2}, and Circa \cite{reluMPC3} have proposed optimizations to optimize MPC CNNs. \cite{wangCharacIspass} characterizes the overheads and pinpoints bottlenecks in MPC inference of Transformer-based models and urges for optimizations for MPC Softmax. MPC-Former~\cite{mpcformer} improves Transformed-based MPC inference by approximating Softmax using ReLUs. Those works propose optimization to ML operators, while MPC-Pipe does not replace any operators and works on system-level optimizations.

\subsection{Other Privacy Preserving Mechanism}
Besides MPC protocols, there exist other families of privacy-preserving mechanisms for ML, such as trusted execution environments~\cite{tramer2019slalom, origami, darkNight, 3leg, fedvault}, differential privacy~\cite{flemings2024differentially,flemings2024ddd,mcmahan2017learning,li2021large,yu2021differentially,edge}, homomorphic encryption~\cite{homoEnc, cpuHEACC2, cpuHEACC3,heAcc1, heAcc2, heAcc3}, coded computing~\cite{yu2019lagrange, tingipdps}, ORAM~\cite{oram, pathoram, ringoram, pageoram, laoram}, and Machine Unlearning~\cite{bourtoule2020machine, guo2023certified, gao2024ethos}. Those works use different threat models and provide different security guarantees.

%% file: 3.background.tex
\textbf{Notations:} We use capitalized letters such as $Z$, $X$, and $Y$ to denote operands in the plaintext, and we use lowercase letters such as $[x]$ and $\langle x \rangle$ to denote secret shares of the original operands.

\subsection{Secret Sharing}
Multiparty computing protocols allow MPC clients to distribute inputs to MPC servers in secret and MPC servers using encrypted inputs to compute the final computation results. Operands are distributed to MPC servers in the format of secret shares, and any secret shares sent to MPC servers should leak no information about the original operand $X$. There are two major ways to share an operand $X$ to $N$ MPC servers: 1) Additive (arithmetic) sharing: An MPC server $i$ will receive $[x_i] : X = \sum_{i=0}^{N-1} [x_i]$ from MPC clients, 2) Binary sharing: An MPC server will receive $\langle x_i \rangle : X = \bigoplus_{i=0}^{N-1} \langle x_i\rangle$ from MPC clients.

$[x_i]$ represents the additive secret share of original operand $X$ in the MPC server $\#i$, and $\langle x_i\rangle$ represents the binary secret share of original operand $X$ in the MPC server $\#i$. In the latter sections, if a subscript is not specified, $[x]$ and $\langle x\rangle$ represent secret shares of operand $X$ sent to some MPC server.

There are many ways to generate secret shares. For example, in the 2PC setting, two additive shares of $X$ can be $[x_0] = X - R$ and $[x_1] = R$, where both $X$ and $R$ are in the same algebraic number field, and $R$ is sampled from a uniform random variable. The uniformity of $R$ renders both $R$ and $X-R$ to leak no information about the original operand $X$. In the same 2PC setting, two binary shares for $X$ can be $\langle x_0\rangle = X \oplus R$ and $\langle x_1 \rangle = R$. Note that MPC-Pipe is agnostic to the additive or binary secret-sharing scheme. Usually, the additive sharing format is more suitable for multiplications and additions, and the binary sharing format is more suitable for bit-wise operations like $XOR$ and shifting. After receiving secret shares from the client, it's the MPC servers' turn to run computations on their local shares. The next several sections will describe protocols to compute multiplications, bit-wise operations, several non-linear operations, and secret share format conversions.

\subsection{Beaver Triple Assisted Operations}
\label{sec:beaverTriple}
For operations such as additions and $XOR$s between $X$ and $Y$, MPC servers just need to add or $XOR$ their own local shares, and then the client can reconstruct the final result as shown below: 
\begin{align}
    \sum_{i=1}^{N} [x_i] + [y_i] &=  \sum_{i=1}^{N} [x_i] + \sum_{i=1}^{N} [y_i] = X + Y\\
    \bigoplus_{i=1}^{N} \langle x_i \rangle \oplus \langle y_i\rangle 
     &= \bigoplus_{i=1}^{N} \langle x_i \rangle \oplus \bigoplus_{i=1}^{N} \langle y_i \rangle = X \oplus Y
\end{align}
However, similar rules do not apply to multiplication and $AND$ operations. In MPC, those operations are assisted by Beaver triples. Algorithm~\ref{alg:beavermult} shows the multiplication assisted with Beaver Triples. Since $AND$ operations for binary shares are equivalent to multiplications for additive shares,  we will only show protocols for multiplication for conciseness. To derive the MPC $AND$ algorithm, one needs to replace every addition with $XOR$ and every multiplication with $AND$ in algorithm~\ref{alg:beavermult}.

\textbf{Beaver Triple Generation in Offline Phase:}
Beaver triples are three numbers in the same numerical field such that $C=A \cdot B$, and the triple is additively shared to MPC servers. $A$ and $B$ are random values. The process of MPC servers deriving shared Beaver Triples is called the offline phase of MPC multiplications because the offline phase is not dependent on any multiplication operands, and it can be performed ahead of any input.  

\textbf{Beaver Triple Usage in Online Phase:}
MPC servers will follow the online phase Algorithm~\ref{alg:beavermult} to compute $X \cdot Y$ using the additive shares of $X$,$Y$, and the Beaver Triples. This paper focuses on online phase runtime reduction, which is input-dependent.

\begin{algorithm}[h]
   \caption{The Online Phase of Beaver Triple Assisted MPC Multiplication}
   \label{alg:beavermult}
\begin{algorithmic}
   \STATE {\bfseries Input:} $[x_i]$, $[y_i]$, $[a_i]$, $[b_i]$ and $[c_i]$ s.t. $C = A \cdot B$

   \STATE {Computes} $[x_i] - [a_i]$ and $[y_i] - [b_i]$
   \STATE {Broadcast} local $[x_i] - [a_i]$ and $[y_i] - [b_i]$
   \STATE Wait until other $[x_i] - [a_i]$ and $[y_i] - [b_i]$ has been received
   \STATE Computes $\Delta = \sum_{i=1}^{N}[x_i] - [a_i]$ 
   \STATE Computes $\epsilon = \sum_{i=1}^{N}[y_i] - [b_i]$ 
   \STATE Party \# 1 computes $[z_1] = [c_1] + \Delta \cdot [b_1] + \epsilon \cdot [a_1] + \Delta \cdot \epsilon$
   \STATE Other parties compute $[z_i] = [c_i] + \Delta \cdot [b_i] + \epsilon \cdot [a_i]$
   \STATE \textbf{Return}: $[z_i]$
\end{algorithmic}
\end{algorithm}

After the online phase, for MPC clients to recover the final product of multiplication, they need to sum all the $[z_i]$s from MPC servers, such that $\sum_{i=1}^{N} [z_i] = XY = Z$.
Note that the Beaver Triples $A$ and $B$ are sampled from a uniform random variable such that $X-A$ and $Y-B$ leak no information about $X$ and $Y$.

\textbf{Comparison Operation with Share Conversion:}
Comparison operations are used in non-linear functions in ML algorithms. For instance, operations such as ReLU and MaxPool need such a comparison. Performing comparisons using MPC is non-trivial. For instance, to compute \textit{Less than}, MPC servers compute $MSB([x_i - y_i])$, where $MSB$ is the function to obtain the most significant bit. Such computations require bit-wise operations. Bit-wise operations like shifting are more efficient in the binary sharing format. Thus, MPC servers need to convert $[x_i - y_i]$ to binary sharing format. Algorithm~\ref{alg:A2B} shows such a conversion process.

\begin{algorithm}[h]
   \caption{Additive Share to Binary Share Conversion}
   \label{alg:A2B}
\begin{algorithmic}
   \STATE {\bfseries Input:} $[x_i]$
   \STATE {Generate} binary shares $\langle [x_i] \rangle_j : [x_i] = \bigoplus_{j=1}^{N} \langle [x_i] \rangle_j$
   \STATE Send $\langle [x_i] \rangle_j$ to party $j$ 
   \STATE Wait until all  $\langle [x_j] \rangle_i$ are received
   
   \STATE Use binary operations to compute $\langle x_i \rangle =  \langle \sum_{j=1}^{N} [x_j] \rangle_i$
   \STATE \textbf{Return} $\langle x_i \rangle$
\end{algorithmic}
\end{algorithm}
The first two lines in algorithm~\ref{alg:A2B} are steps for each MPC server to share its $[x_i]$ to other MPC servers in binary sharing format. Upon completion of this step, each MPC server will have a binary share for every additive share of the original operand $X$. In the last step of algorithm~\ref{alg:A2B}, each MPC server will need to perform a series of $AND$ and $XOR$ operations (bit-wise logical only) to obtain the summation of binary shares. Such logical operations, using the Set-Propagate-Kill (SPK) tree, can be found in hardware adders~\cite{weste2015cmos}. In the MPC framework on which our work is based, ReLUs and Maximum functions all use algorithm~\ref{alg:A2B} to perform conversions to obtain the most significant bit. The SPK tree is described further in a later section as well.


\subsection{The number of parties}
The security guarantee of MPC is measured by the number of MPC servers involved. In our semi-honest majority setting, for an adversary to reconstruct the original secret $X$, the adversary has to compromise every MPC server. Even if the adversary compromised $N-1$ MPC servers, the retrieved data would appear random to the adversary. Luckily, SS-based MPC allows an arbitrary number of parties to participate in the MPC protocol. Thus, the more MPC parties involved, the stronger the security guarantee is. However, a stronger security guarantee comes with a cost. Revealing of $X-A$ and $Y-B$ in Beaver Triple-assisted protocol (as described in Algorithm~\ref{alg:beavermult}) become more expensive as we increase $N$.

%% file: 4.mpcpipe.tex
In this section, we will describe the three major techniques in MPC-Pipe: 1) inter-linear pipeline, 2) inner-layer pipeline, and 3) inter-batch pipeline. We will also discuss how those techniques apply to different general Machine Learning layers. The first two schemes aim to remove unneeded computation and communication data dependencies within ML layers during training or inference, while the last scheme aims to overlap communication and computation between different batches. After removing false data dependencies between communication and computation, MPC-Pipe can improve both the latency and the throughput of ML training and inference.

\subsection{Inter-linear pipeline for inference and training}
The first scheme focuses on overlapping computation and communication in linear layers in ML models: Dense and Convolution. We will describe the inter-linear pipeline using the Convolution layer of an inference/forward pass as an example. We then discuss how the inter-linear pipeline can be used to optimize training as well. Our goal is to compute $Z = Conv(X, W)$ without revealing $W$ or $X$ using MPC, where $W$ is the weight of the layer, and $X$ is the input.
MPC servers need to secret share a Beaver Triple $A$, $B$, and $C$ such that $C = Conv(A, B)$. Dimensions of $A$ and $B$ will be the same as $X$ and $W$. Similar to the Algorithm~\ref{alg:beavermult} from Section~\ref{sec:beaverTriple}, MPC servers broadcast $[x-a]$ and $[w-b]$ before performing the computationally heavy operation of
\begin{align}
    \label{eq:convz}
    [z] = &[c] + Conv((X-A), [b]) + Conv([a], (W-B)) \nonumber \\
     &+ Conv((X-A), (W-B))
\end{align}

\begin{figure}[h]
  \centering
    \includegraphics[width=7cm]{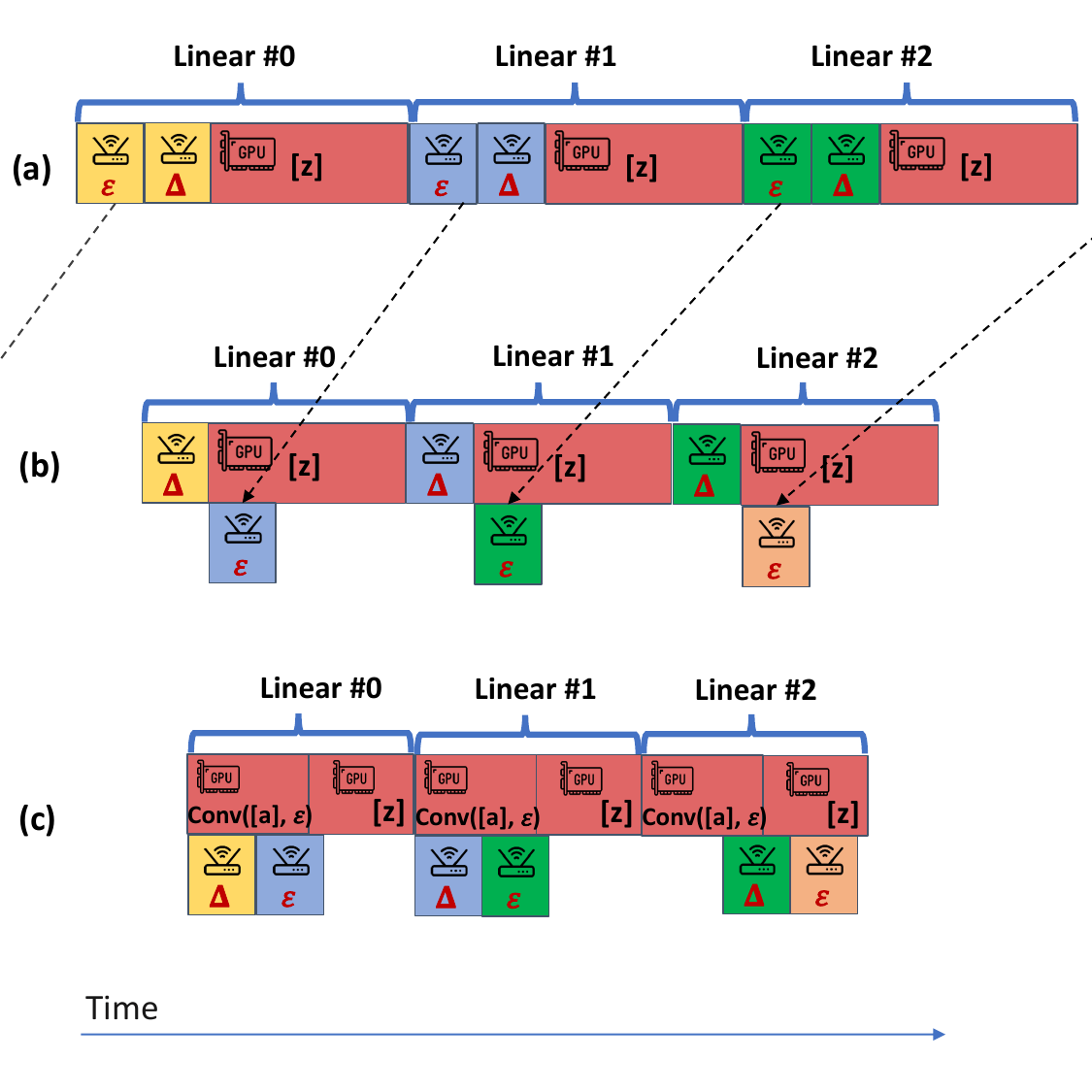}
  \caption{Inter-linear pipeline demonstration; the box with a GPU icon is the time spent for computation; the box with a router icon is the time spent transmitting data.}
  \label{fig:inter-pipe}
\end{figure}

Note that $(X-A)$ (referred to as $\Delta$ value in section~\ref{sec:beaverTriple}) is first computed by summing all the  $[x_i - a_i]$  received as broadcasts from all other parties. Similarly, $(W-B)$ (referred to as $\epsilon$) is computed by summing all the  $[w_i - b_i]$  received as broadcasts from all other parties. It is thus intuitive to expect that the two MPC servers must wait until the broadcast is complete before they proceed to compute Equation~\ref{eq:convz}. The computation of $[z]$ depends on $X-A$ and $W-B$ as shown visually in Figure~\ref{fig:inter-pipe}(a), where $\epsilon=W-B$ and $\Delta=X-A$.

\textbf{Key Idea 1: Transmit input independent $[w - b]$ asynchronously. } 
During forward propagation, the computation of $W-B$ is only related to the Convolution weight $W$ and the Beaver triple value $B$, neither of which depends on the input $X$. As $W$ and $B$ are available well before the start of the forward pass, each party can compute and broadcast $[w - b]$ asynchronously. Therefore, each party can calculate $[w^{next} - b^{next}]$ for any upcoming layer $l_{next}$ and broadcast this value before the computation of $[z]$ for that layer begins.

For instance, as illustrated in Figure~\ref{fig:inter-pipe}(b), linear layer $\#1$ can start broadcasting $[w^{l2}-b^{l2}]$ for linear layer $\#2$ ($l2)$ before the computation of $[z]$ for linear layer $\#1$. Thus while computing $[z]$ for the current layer $\#1$, the MPC servers concurrently transmit  $[w^{l2}-b^{l2}]$ for layer $\#2$. Note that the MPC servers would have already received $[w^{l1}-b^{l1}]$ for linear layer $\#1$ during the execution of the convolution computation for linear layer $\#0$. 


\textbf{Key Idea 2: Overlap input dependent $[x - a]$ transmission with  Conv operation.} 
For batched execution with multiple inputs, the size of $[x-a]$ could be bigger than that of $[w-b]$. Hence, MPC-Pipe creates additional opportunities for computation and communication overlap. MPC-Pipe can also hide the latency of transmitting $[x-a]$,  besides hiding the latency of sending $[w-b]$. In particular, the MPC server issues the broadcast of $[x-a]$ in a non-blocking manner. While the broadcast runs in the background, the MPC-Pipe initiates the  $Conv([a], (W-B))$ computation.  The overlap of the transmission of $[x-a]$ with convolution computations is visually illustrated in Figure~\ref{fig:inter-pipe}(c). Using the inter-linear pipeline scheme of MPC-Pipe, the new algorithm for computing Convolution with MPC is shown in Algorithm~\ref{alg:linear-pipe}. Note that the secret shares with a $next$ symbol, such as $[w_i]^{next}$ and $[a_i]^{next}$, are secret shares needed by a later linear layer. By efficiently pipelining the inter-linear dependencies, the wait time in lines 4 and 7 of Algorithm~\ref{alg:linear-pipe} is minimized. In fact, as we demonstrate later in the results, by overlapping computation with the transmission of both $[x-a]$ and $[w-b]$, MPC-Pipe is able to hide all the communication overheads associated with the linear layers. Note that in our implementation, we only needed to transmit one layer ahead to hide the communication delays during forward propagation. However, one may choose to transmit data for any future layer if the communication delays are much longer.  

\textbf{Application to the backward pass:} We have discussed how the inter-layer pipeline can be applied to the inference/forward pass. The inter-linear pipeline can also be applied to ML's backward pass. In the forward pass, MPC parties compute Convolution between $X$ and $W$, while in the backward pass, MPC parties need to compute gradients w.r.t. the weights and input. To compute gradients w.r.t. the weights, MPC parties need to compute the convolution between output gradients and the intermediate activation (computed during the forward pass). To compute gradients w.r.t. the input, MPC parties need to compute convolution between the output gradients and the weight. Before the start of the backward pass, both the intermediate activation and the weight of the linear layers are already available. Thus, we can directly apply the inter-linear pipeline to hide data transmission required by convolutions in the backward pass for training. Thus, our inter-linear pipeline approach can be applied to both training and inference workloads. 

\begin{algorithm}[h]
   \caption{The Online Phase of MPC Convolution w/ Inter-linear pipeline for party $i$.}
   \label{alg:linear-pipe}
\begin{algorithmic}[1]
    \STATE {\bfseries Input:} $[x_i]$, $[w_i]$, $[w_i]^{next}$, $[a_i]$, $[b_i]$, $[b_i]^{next}$ and $[c_i]$ s.t. $C = Conv(A, B)$;
    \STATE {Computes} $[x_i] - [a_i]$ and $[w_i]^{next} - [b_i]^{next}$;
    \STATE {Initiate the broadcasting of} local $[x_i] - [a_i]$ and $[w_i]^{next} - [b_i]^{next}$;
    \STATE Wait until all other $[w_i] - [b_i]$ (initiated in a previous linear layer) have been received;
    \STATE Compute $\epsilon = W - B = \sum_{i=1}^{N}[w_i] - [b_i]$ ; 
    \STATE Compute $[z]_i = Conv([a_i], \epsilon)$;
    \STATE Wait until all other $[x_i] - [a_i]$ has been received;
    \STATE Compute $\Delta = X - A =  \sum_{i=1}^{N}[x_i] - [a_i]$; 
    \STATE Party \# 1 computes $[z_1] += [c_1] + Conv(\Delta,[b_1])) + Conv(\Delta, \epsilon)$;
    \STATE Other parties compute $[z_i] += [c_i] + Conv((X-A),[b_i]))$;
    \STATE \textbf{Return}: $[z_i]$;
\end{algorithmic}
\end{algorithm}

\subsection{Inner-layer pipeline}
The previous section enabled communication and computation phases to be overlapped during linear operations. There exists a class of non-linear layers in many ML models that do not have learned weights. Those layers include ReLU, Softmax, and MaxPooling. These layers use the activations generated from prior linear layers where the data is already in arithmetic secret sharing format. Without static weights to use, these non-linear layers can not exploit the inter-linear pipeline scheme. 

As discussed in the background, the state-of-art semi-honest MPC protocols use conversion algorithms, such as those developed in ~\cite{aby,crypten2020},  to convert arithmetic secret shares into binary secret shares to perform comparisons needed in ReLU, Softmax, and MaxPooling. Each comparison will require one conversion from arithmetic to binary. For example, ReLU requires one conversion, and MaxPooling requires two conversions if the selected filter size is $2\times2$ (and 4 conversions for $4\times4$). These conversions dominate the total latency of performing non-linear computations in the MPC setting, as has also been shown in prior work~\cite{wangCharacIspass}.

{The conversion algorithm is described in Algorithm~\ref{alg:A2B}. As discussed in the background, the last step in the Algorithm~\ref{alg:A2B} performs addition among $\langle [x_j] \rangle_i$. To sum $\langle [x_j] \rangle_i$, which are binary secret shares, MPC parties will need to execute Set-Propagate-Kill (SPK) Tree loops.} The SPK tree loop involves iteratively evaluating six beaver triple-assisted $AND$ operations (line 14 in the loop) and several other local logical operations (lines 9-12). SPK tree circuit logic is shown:
\begin{lstlisting}[language=Python]
def SPK_circuit(S, P):
    mask, out_masks, mult = SPK_constants()
    S1P1 = stack(S, P)
    for i in range(6):
        in_mask = mask[i]                
        out_mask = out_masks[i]          

        # constants
        not_out_mask = out_mask ^ -1       
        P0 = SP[1] & out_mask               
        S1P1 = SP & in_mask                
        S1P1 *= mult[i]   
        
        update = P0 & S1P1  # beaver AND
        SP[1] = SP[1] & not_out_mask # const AND
        SP ^= update
    return SP[0], SP[1]
\end{lstlisting}

\begin{figure}[h]
  \centering
      \includegraphics[width=7cm]{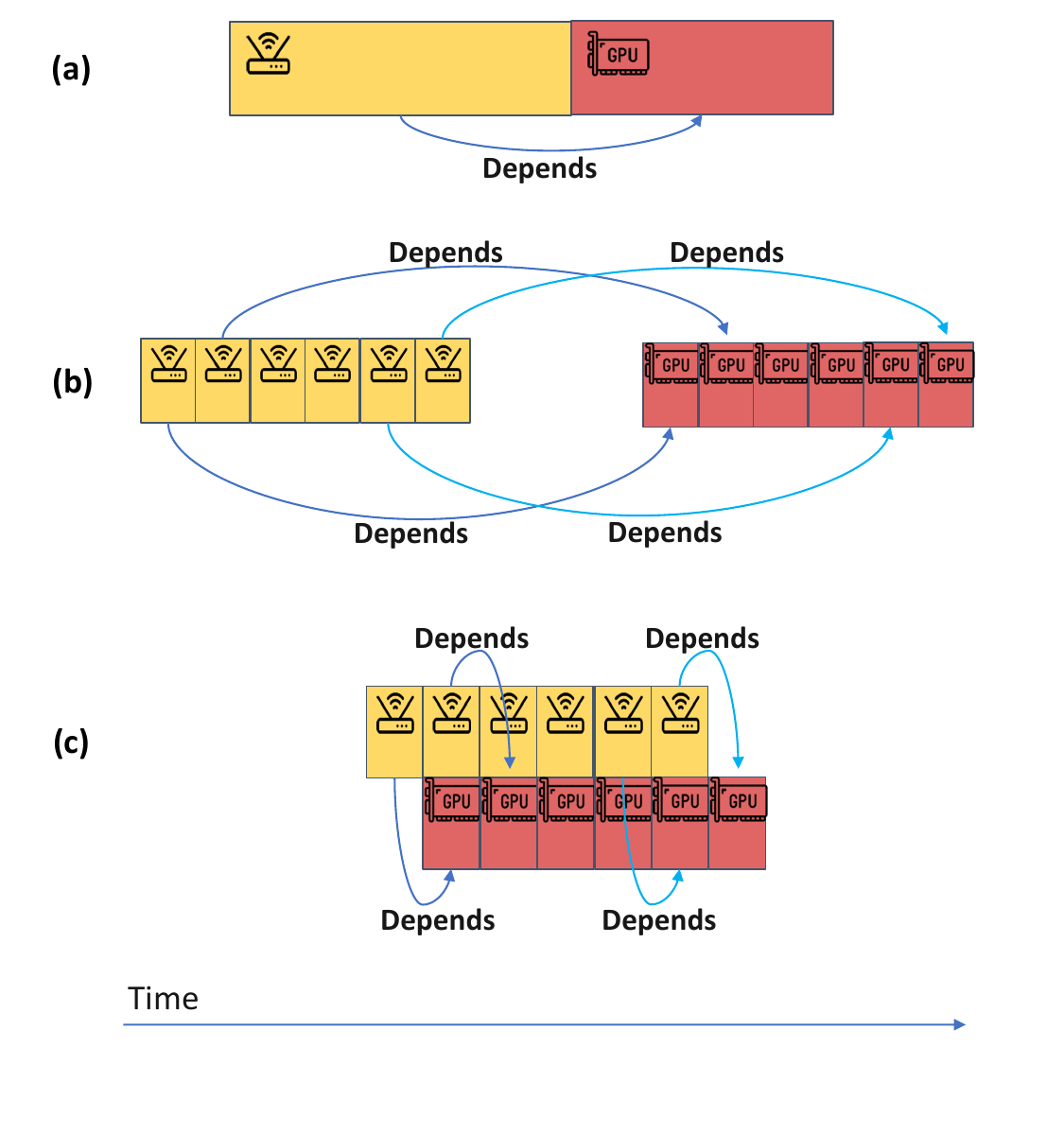}
  \caption{Inner-layer pipeline demonstration; the box with a GPU icon is the time spent for computation; the box with a router icon is the time spent transmitting data.}
  \label{fig:inner-pipe}
\end{figure}

Given that the SPK tree has multiple AND (line 14) operations, these operations are assisted by Beaver triples and will follow an algorithm similar to Algorithm~\ref{alg:beavermult}. To perform $AND$ operations, MPC servers need to broadcast $\langle x \oplus a \rangle$ and $\langle y \oplus b\rangle$ before computing the final 
\begin{align}
\label{eq:andComp}
\langle z \rangle = \langle c \rangle \oplus ((X\oplus A) \& \langle b \rangle)  \oplus (\langle a \rangle \& (Y \oplus B))
\end{align}
due to data dependencies as shown in Figure~\ref{fig:inner-pipe}(a) (the yellow box collectively shows transmission of $\langle x \oplus a \rangle$ and $\langle y \oplus b\rangle$).

In the above code, the variables ``mask'', ``out\_mask'', and ``mult'' returned by the function \textit{SPK\_constants} are public constants, so the element-wise binary operations in lines 9-12 do not need a Beaver Triple to assist their computation. Thus, the AND and the preceding logical operations are all local element-wise operations on secret shares. 

\textbf{Key Idea: Tiling Element-wise Operations} We make the key observation that the six $AND$ operations and the preceding logical operations in the SPK tree are element-wise operations. Hence, there is no need to wait for the entire input vector computation to be completed before transmitting the vector. The data dependency graph shown in Figure~\ref{fig:inner-pipe}(a) is unnecessary. For ML models with a large number of parameters, this coarse-grain transmission is particularly inefficient since the size of the activation maps are relatively large.  MPC-Pipe breaks the large input to the SPK tree computation into several smaller tiles. The granularity of the data dependency graph transforms into a finer graph, as shown in Figure~\ref{fig:inner-pipe}(b). This MPC-Pipe optimization removes the large granularity communication dependency of AND and other logical operations by breaking inputs into many smaller groups. Thus, MPC servers will initiate the transmission process for group $\# i+1$ before performing computation for group $\# i$. In this way, group $\# i$ does not have to wait for metadata that is not related to its computation. We call this approach \textit{inner-layer pipeline}.

Using the inner-layer pipeline, MPC servers can proceed to compute on one part of the larger input without waiting for the complete transmission of all the input data. Figure~\ref{fig:inner-pipe}(c) illustrates an idealized timing diagram of the inner-layer pipeline where compute and communication times are the same. In our experimental setting, the compute time per item of data during format conversion tends to be smaller than the communication time. Hence, in our setting, we achieve a partial overlap of communication with computation. Algorithm~\ref{alg:inner-pipe} shows the MPC AND algorithm with the inner-layer pipeline.

\textbf{Rounds of communications:} One might argue that breaking the transmission of a large input into multiple smaller transmissions will increase the number of rounds of communications and result in more propagation delay. This reasoning is only true when the $M$ transmissions in line 5 of Algorithm~\ref{alg:inner-pipe} are synchronous or separated by a large interval (larger than the propagation delay in the WAN). In our system implementation, transmissions in line 5 are asynchronous, and the computations in line 4 always take less time than the transmissions. Thus, transmissions in line 5 are started back-to-back and share a single propagation delay. Consequently, in the setting with a large network propagation delay, the inner-layer pipeline does not introduce more propagation delay. 

\textbf{Application to the backward pass:} The operations in the backward pass of the non-linear layer usually involve many element-wise multiplications. For example, the forward pass of ReLU requires MPC parties to perform expensive comparison, while the backward pass of ReLU only requires an element-wise multiplication between the output gradients and an input mask. The input mask's size is computed during the forward pass, whose size is the same as the output gradient that is secretly shared. Since those operations are also element-wise operations, the inner-layer pipeline that breaks false dependencies in long vectors can be applied to the training process as well.

\begin{figure}[h]
  \centering
      \includegraphics[width=5cm]{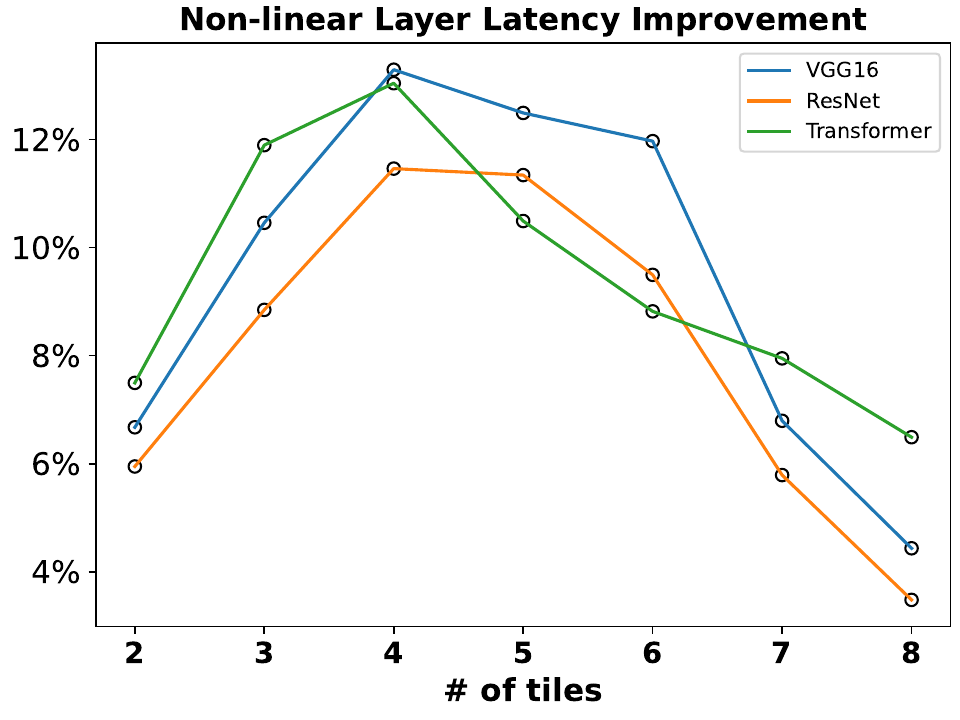}
  \caption{Latency improvement for non-linear layers with different numbers of tiles.}
  \label{fig:abalation}
\end{figure}

\revision{\textbf{The number of tiles:} We experimentally measure the runtime improvement of inner-layer pipelines using different numbers of tiles. On average, when using four tiles, inner-layer pipelines perform best.   Figure~\ref{fig:abalation} shows the latency improvement for non-linear layers with different numbers of tiles. As the number of tiles grow past 4 the size of each tile shrinks and the parallel computation efficiency to perform element-wise operations on GPUs reduces. Hence, there is a tradeoff between the total number of elements to operate, the size of each tile and GPU efficiency. In general we recommend keeping the tile size large enough so as to maximize the GPU usage for element-wise operations.}

\begin{algorithm}[h]
   \caption{The Online Phase of MPC AND w/ Inner-layer pipeline for party $i$.}
   \label{alg:inner-pipe}
\begin{algorithmic}[1]
    \STATE {\bfseries Input:} $[x_i]$, $[y_i]$, $[a_i]$, $[b_i]$, and $[c_i]$ s.t. $C = A\ \&\ B$;
    \STATE Split $[x_i]$, $[y_i]$, $[a_i]$, $[b_i]$, and $[c_i]$ into $M$ batches;    
    \FOR{m in $0\ ..\ M - 1$}
        \STATE {Computes} $[x_i]_m \oplus [a_i]_m$ and $[y_i]_m \oplus [b_i]_m$;
        \STATE {Initiate the broadcasting of} $[x_i]_m \oplus [a_i]_m$ and $[y_i]_m \oplus [b_i]_m$;
    \ENDFOR
    \FOR{m in $0\ ..\ M - 1$}
        \STATE Wait until all other $[x_i]_m \oplus [a_i]_m$ and $[y_i]_m \oplus [b_i]_m$ have been received;
        \STATE Compute $\Delta =  X \oplus A =  \sum_{i=1}^{N} [x_i]_m \oplus [a_i]_m$
        \STATE Compute $\epsilon = Y \oplus B = \sum_{i=1}^{N} [y_i]_m \oplus [b_i]_m$
        \STATE Party \# 1 computes $[z_1]_m = [c_1]_m \oplus (\Delta \ \&\ [b_1]_m) \oplus (\epsilon\ \&\ [a_1]_m) \oplus (\Delta \ \&\ \epsilon)$;
        \STATE Other parties compute: $[z_i]_m = [c_i]_m \oplus (\Delta \ \&\ [b_i]_m) \oplus (\epsilon \ \&\ [a_i]_m)$;
    \ENDFOR
    \STATE $[z_i] = Concat([z_i]_0, [z_i]_1, ... , [z_i]_{M-1})$;
    \STATE \textbf{Return}: $[z_i]$;
\end{algorithmic}
\end{algorithm}


\begin{figure}[h]
  \centering
      \includegraphics[width=7cm]{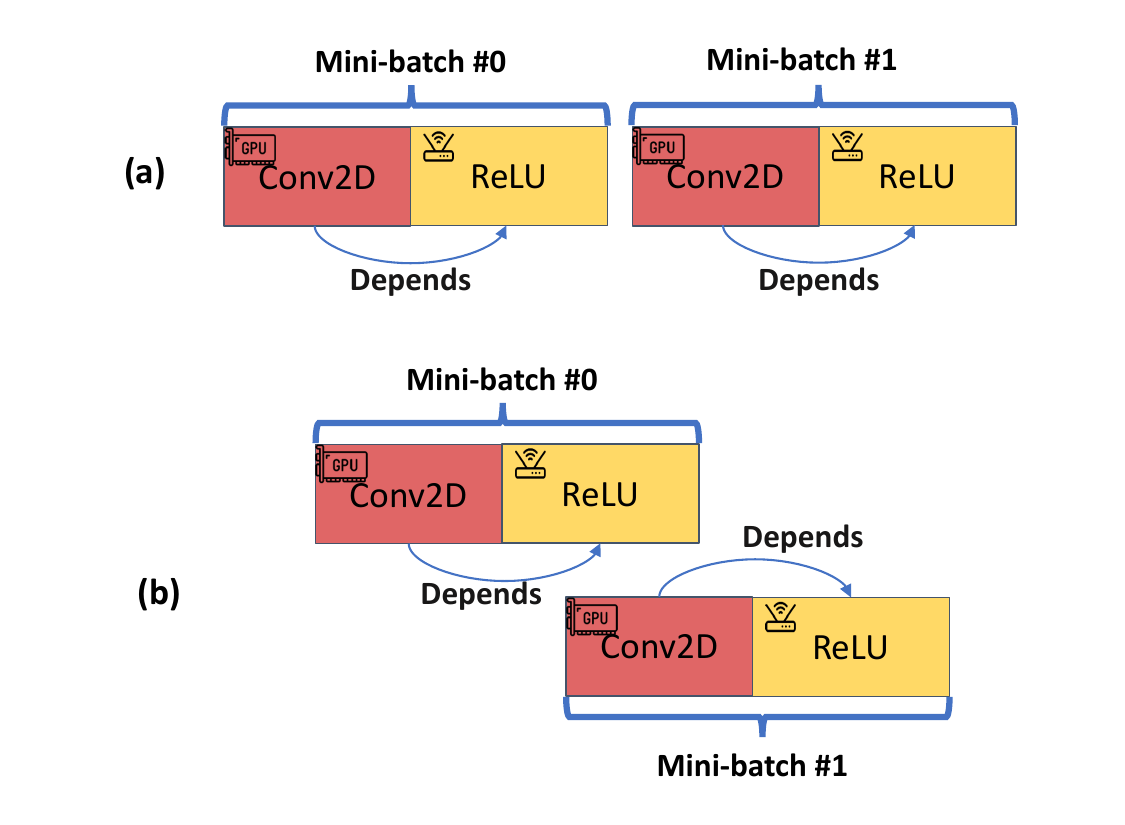}
  \caption{Inner-batch pipeline demonstration; the box with a GPU indicates that operation is dominated by computation; the box with a router icon indicates that operation is dominated by communication.}
  \label{fig:inter-batch}
  \vspace{-4mm}
\end{figure}

\subsection{Inter-batch pipeline}
The first two techniques aim to remove false data dependencies and overlap the computation and communication in an inference or training run. 
Additional opportunities exist in a batched training or inference mode. Multiple input mini-batches allow us to overlap the computation of the Convolution layer for mini-batch $\#2$ and communication of the ReLU layer for mini-batch $\#1$ as illustrated in Figure~\ref{fig:inter-batch}. In MPC-Pipe, we implemented this batch pipelining using two threads in each MPC server; one thread continuously executes linear layers, and another thread continuously executes nonlinear layers. Two mini-batches are fed into the model. The linear thread will evaluate the first linear layer of the model using the first mini-batch and send the result to the corresponding nonlinear layer. Instead of waiting for the nonlinear thread to return the results, the linear thread will evaluate the first linear layer again using the second mini-batch, overlapping the computation of the Convolution layer for mini-batch $\#2$ and communication of ReLU layer for mini-batch $\#1$.

\subsection{Impact on latency and throughput}
Latency is the time taken for a model to produce a result, which is the summation of the latency of all layers. The first pipeline scheme aims to reduce the latency of linear layers, and the second pipeline scheme aims to reduce the latency of nonlinear layers. Those two techniques combined can reduce end-to-end training and inference latency. With shorter latency, those two techniques can improve the throughput as well. On the contrary, the third technique utilizes layer-wise parallelism that achieves a higher degree of computation and communication overlapping between two adjacent mini-batches. Thus, the third technique produces better throughput but does not reduce latency. In our experimental section, when applying all the pipeline schemes, MPC-Pipe improves both the latency and throughput, indicating the importance of all three schemes working in unison. 

%% file: 5.evaluation.tex
\subsection{Experimental Setups}

\revision{
In our experiment, each MPC party is a server with the following configurations: CPU: Intel Xeon Gold 5220R CPU 24 cores, OS: Ubuntu 20.04.6, Nvidia Quadro RTX 5000, CPU DRAM: 512 GB, Network card: InfiniBand CX353A(40Gb/s). 
We use software “wondershaper” and “iproute2” to simulate slower network and network latency, and there are no workloads other than MPC ML running on each MPC party. When we say interconnection bandwidth is 10Gbps, we mean that each MPC party will have an upload/download bandwidth of 10Gbps.}

Implementation of MPC-Pipe is based on CrypTen~\cite{crypten2020}, an MPC framework based on PyTorch from Meta AI. We gather inference and training runtime from the average of \textbf{50} iterations so as to remove the influence of random network fluctuations. \revision{The inference and training runtime includes input distribution and online inference and training circuit evaluation. Note that in our experimental setup, the input distribution time is measured to be less than $0.03\%$ of the total execution runtime.} All the results presented are end-to-end system throughput and latency improvements on existing hardware without requiring MPC protocol changes or weakening the threat model.   


\subsection{Enhanced CrypTen Baseline}
We choose a modified CrypTen~\cite{crypten2020} as our baseline.  CrypTen is based on PyTorch, and it supports efficient MPC operations on GPUs. Given the wide use of GPUs, particularly in training, we focus on implementing MPC-Pipe on GPUs. 
Other MPC frameworks ~\cite{kumar2020cryptflow, delphi, mp-spdz, gazzele, chameleon, secureml, fis} either do not use semi-honest security model or do not fully support GPUs. Thus, we choose CrypTen for its full support on GPUs and its support of the semi-honest threat model.

The original CrypTen does not separate the offline (Beaver Triple generation) and the online phase. However, the real-time performance of MPC is not dependent on the offline phase since the Beaver Triple generation is done out of the critical path.  We thus modified CrypTen to separate the offline phases from the online phases. This modified CrypTen that decoupled the offline and online phases is used as the baseline in our experiments. We then implemented all the pipeline schemes on top of the modified Crypten framework to measure the performance of the MPC-Pipe. 



\subsection{Models evaluated}
We evaluated MPC-Pipe on two major Machine Learning workloads: 1) a Transformer encoder and 2) Convolution Neural Networks. \revision{The Transformer encoder is a single layer Transformer encoder with a configuration of (H=1024, A=16), and the input sequence length is 512, which is the identical Transformer configuration to BERT-large and XLM-R. CNN models we evaluated are ResNet-50 (25m params, ImageNet size input) and VGG16 (138m params, ImageNet size input)}.


\input{charts/throughput_chart}

\subsection{Throughput}
In this section, we will demonstrate how MPC-Pipe can enhance the throughput performance of Machine Learning inference and training. Throughput refers to the number of inputs that can be processed by an ML model within a unit of time (inputs/minute).  \revision{Given $T_{base}$ and $T_{pipe}$ are time in minutes to process $n$ inputs for the baseline and MPC-Pipe, we use $\frac{\frac{n}{T_{pipe}}}{\frac{n}{T_{base}}} - 1 = \frac{T_{base}}{T_{pipe}} - 1$ to calculate the throughput improvement.} Figure~\ref{fig:throuputs} illustrates the throughput enhancements achieved by MPC-Pipe, with a connection bandwidth of 10Gbps, over the enhanced CrypTen baseline. Figures in the first row show the throughput improvements for the inference, and those in the second row show the improvement for training. Each figure shows inference and training improvements in both LAN and WAN (70ms network latency) settings. \revision{Each bar also includes an error bar (one standard deviation) that shows the run-to-run variation across 50 different inference or training runs. In each subfigure, bars marked with ``Pipe$\#1$'' show the throughput improvement when using inter-linear pipeline; bars marked with ``Pipe$\#2$'' show the throughput improvement when using inner-layer pipeline; bars marked with ``Pipe$\#3$'' show the throughput improvement when using inter-batch pipeline; and bars marked with ``combined'' show the throughput improvement when all pipelines above are used.} 
\sheperd{Because MPC-Pipe overlaps compute and communication runtime, MPC-Pipe's throughput improvement is determined by how computation and communication are skewed. For instance, when computation is dominant, then the maximum overlap is limited by the fraction of communication time and vice-versa. We present a more detailed analysis of this breakdown below.}

\input{charts/throuput_categories}


\begin{figure}[h]
  \centering
  \includegraphics[width=8cm]{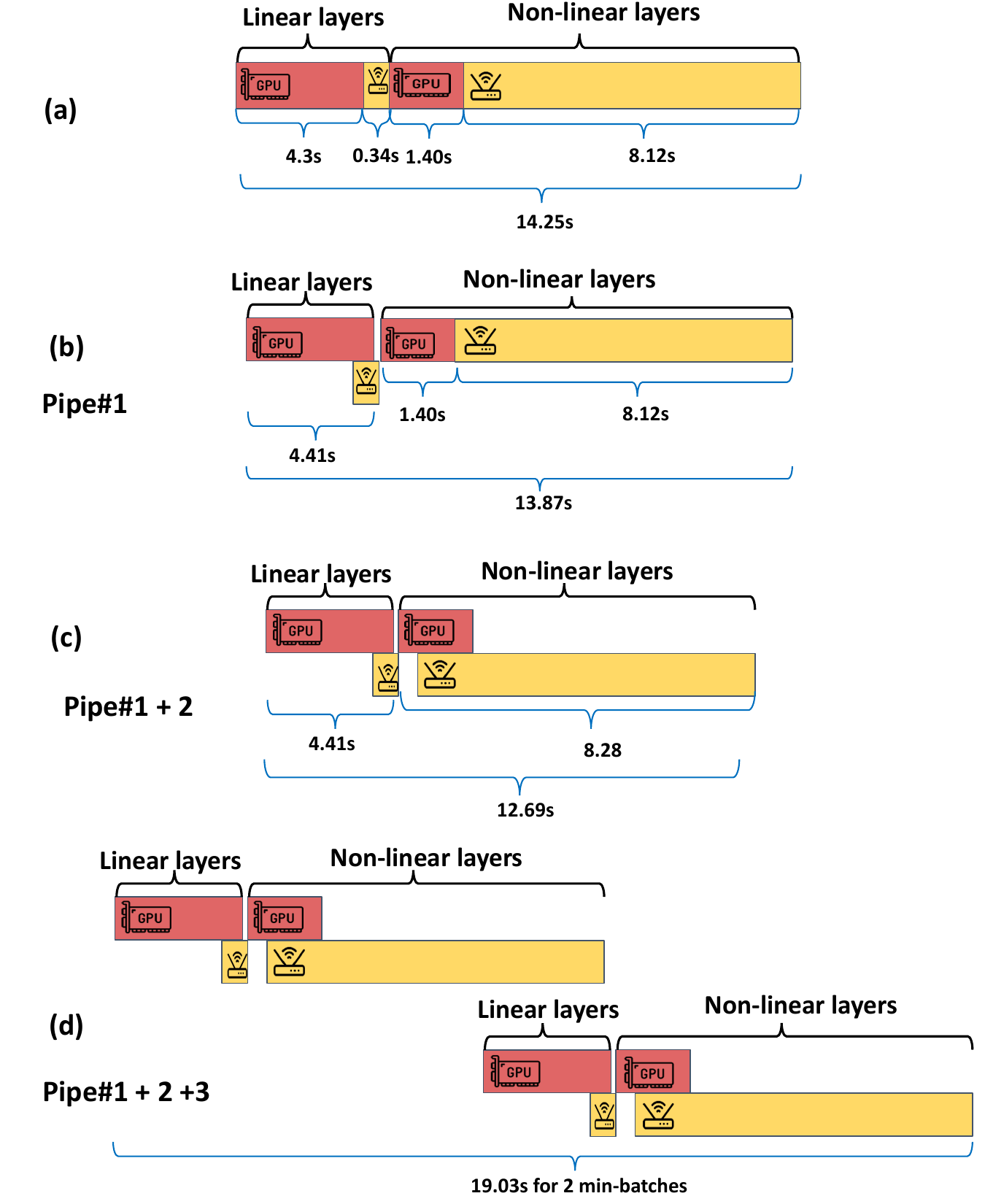}
  \caption{Illustration of MPC-Pipe's impact on VGG16 2PC inference; the box with a GPU icon is the time spent for computation; the box with a router icon is the time spent transmitting data.}
  \label{fig:case-infer}

\end{figure}

\begin{figure}[h]
  \centering
  \includegraphics[width=8cm]{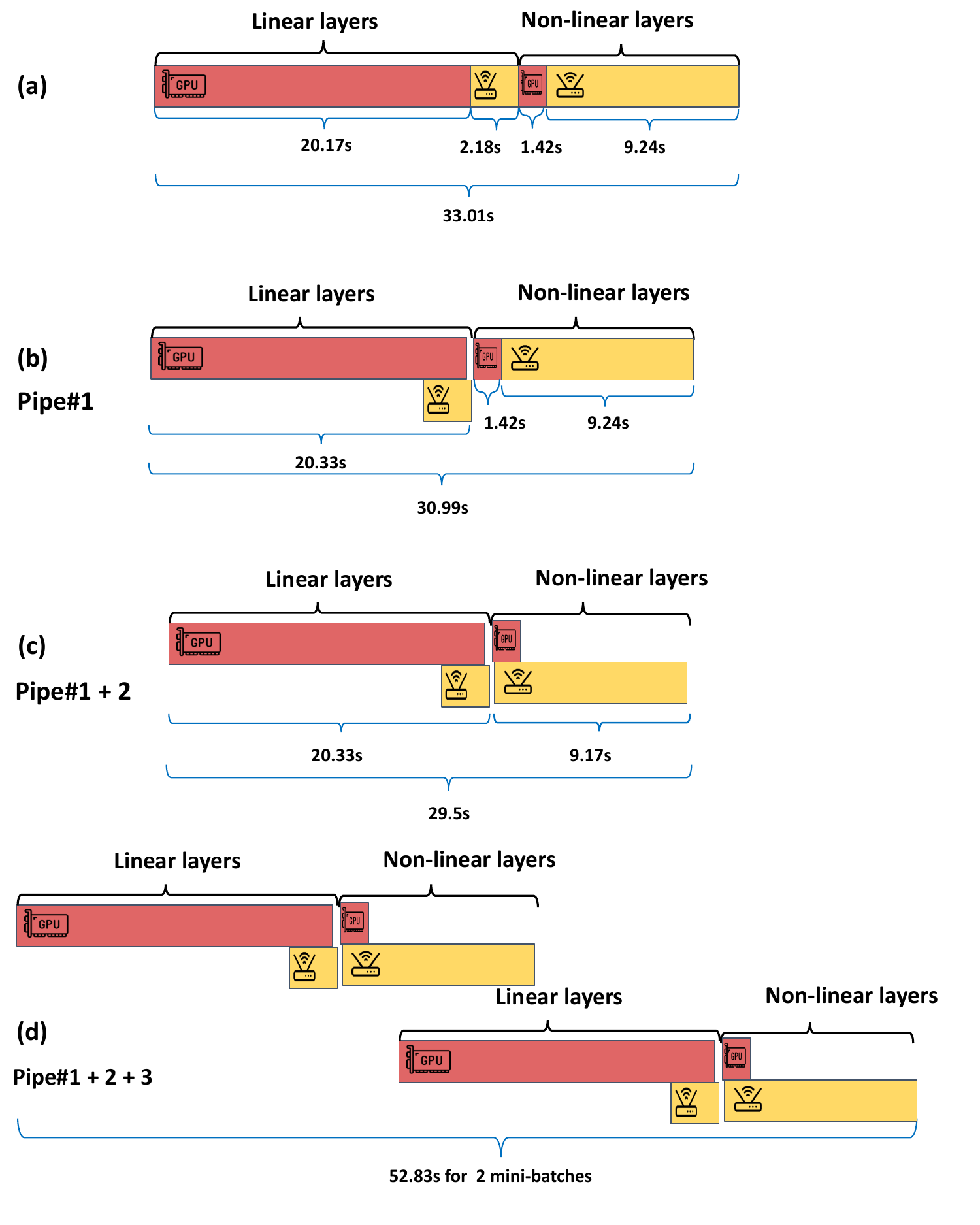}
  \caption{Illustration of MPC-Pipe's impact on VGG16 2PC training.}
  \label{fig:case}
\end{figure}

\subsubsection{Inference}
\revision{Figures in the top row of Figure~\ref{fig:throuputs} demonstrate the throughput improvement achieved by MPC-Pipe for inference. For inference, linear layers are dominated by matrix operations, and the non-linear layers are dominated by data transmission required by secret sharing conversion. Recall that operations such as ReLU and Softmax require additive shares to be transformed into binary shares, and this share conversion requires multiple communication rounds. 
(Section~\ref{sec:background}). As described in Section~\ref{sec:pipe}, linear layers mostly benefit from the inter-linear pipeline (pipe\#1), and non-linear layers mostly benefit from the inner-layer pipeline (pipe\#2). With all three pipeline schemes, MPC-Pipe can improve inference throughput by $\pcent{49.8}$, $\pcent{34.0}$, and $\pcent{50.4}$ for VGG16, ResNet and Transformer, respectively.}

\revision {To provide a concrete and exact measured time, Figure~\ref{fig:case-infer} visually shows how MPC-Pipe's pipeline schemes impact the VGG16 inference.} Figure~\ref{fig:case-infer}(a) shows the breakdown of MPC-based VGG16 inference time, without any pipelining, across linear and non-linear layers. MPC parties spend $14.25$s on VGG16 inference. About 4.64s is spent on linear layers, of which 0.34s is spent on communication. Figure~\ref{fig:case-infer}(b) shows the impact of inter-linear pipelining where the communication latency during linear operations is almost entirely hidden under the linear layer computations. Thus, the 0.34s communication delay is overlapped with computations. Note that due to some initial setup and epilogue, about 0.053s of communication could not be overlapped. Figure~\ref{fig:case-infer}(c) shows the impact of applying the additional inner-layer pipelining where the element-wise operations are tiled and communicated during non-linear operations. However, non-linear layers are dominated by multiple rounds of communication due to share conversion overheads, and hence, inner-layer pipelining is only able to overlap computation under the shadow of the communication. By applying the two pipeline schemes, the overall latency is reduced from 14.25s to 12.69s. Finally, Figure~\ref{fig:case-infer}(d) demonstrates the benefit of inter-batch pipelining where the GPUs that are idle due to non-linear communication bottlenecks in one mini-batch are re-purposed to start processing linear layer operations of another min-batch. Thus, 2 mini-batches were processed in 19.03s, which translates into a throughput of 6.3 images/minute, which is a substantial improvement over 4.21 images/minute for the baseline. The improvements are held for both the LAN and WAN settings, but they are slightly reduced in the WAN setting where the communication delays are more prominent; some of these WAN delays could not be overlapped entirely, leading to slightly reduced throughput.

\subsubsection{Training}
\revision{The bottom row of Figure~\ref{fig:throuputs} demonstrates the throughput improvement achieved by MPC-Pipe for training, which includes forward pass and backward pass. Recall that the backward pass is primarily dominated by the computation of linear layers. For linear layers, backward propagation involves matrix multiplication, such as multiplying output gradients with both the input and weight matrices ($x^T \cdot grad_{out}$ and $grad_{out} \cdot w^T$). Backward operations for non-linear layers like ReLU and MaxPooling involve element-wise multiplication, which is relatively cheap to compute without expensive format conversion. For example, the ReLU backward function will be element-wise multiplication between a secret shared input mask and output gradients. The input mask indicates which elements in the original inputs are non-zero. To summarize, even though ReLU and SoftMax operators in the backward propagation may be non-linear operations, they don't need format conversions. The format conversions are done during forward propagation and are reused during backpropagation. Thus the backpropagation is dominated by matrix multiplications.} 

\revision{Figure~\ref{fig:case} illustrates how each pipeline scheme impacts VGG16's 2PC training. As described in detail for inference, each pipeline scheme individually contributes to latency reduction, and cumulatively, MPC-Pipe is able to process 2 images in 55.82s (2.15 images/minute), while the baseline is only able to process 1.81 images/minute.}

 \sheperd{While Figure~\ref{fig:throuputs} shows MPC-Pipe's throughput improvement on the entire online phase, Table~\ref{tab:throu-cate} summarizes how MPC-Pipe improves the throughput of linear and non-linear layers. Linear layers mostly benefit from the inter-linear pipeline, and non-linear layers benefit from the inner-linear pipeline.}

\subsection{Resource Utilization Improvement}
\revision{One of the additional benefits of MPC-Pipe,  related to the throughput improvements, is the improved resource utilization. By overlapping compute and communication MPC-Pipe is expected to improve the GPU utilization as well as network bandwidth utilization. In our setup, the communication links are $10$Gbps. For baseline VGG16 2PC inference, the average bandwidth observed was $5.91$Gbps. But when applying MPC-Pipe with all three pipelines combined the network utilization increased to $8.85$Gbps.  A similar observation is made for ResNet and Transformer, and the average network utilization is improved from $\pcent{70.73}$ to $\pcent{92.23}$ for Resnet, and from $\pcent{52.3}$ to $\pcent{67.5}$ for Transformer. On the GPU front, MPC-Pipe improves GPU utilization as well. When using baseline protocols for VGG16 2PC, GPUs are idle $\pcent{60}$ and $\pcent{35}$ for inference and training. MPC-Pipe is able to reduce the idle time to  $\pcent{39.5}$ and $\pcent{18.27}$ for inference and training. A similar observation is made for ResNet and Transformer as well.}

\input{charts/latency_chart}

\subsection{Latency}
\label{sec:exp-latency}
This section will showcase MPC-Pipe's latency improvements for inference and training. Latency is defined as the time taken from a model receiving an input to the model producing the final result for that input. 
We break down the computation of CNN models into the following categories: 1) Linear, 2) ReLU, 3) MaxPool, and 4) Softmax (SMax). The Linear category includes Convolution layers and Dense layers. For VGG16 and ResNet, most of the Linear layers are Convolution, and for Transformer models, all the Linear layers are matrix multiplications.

\subsubsection{Inference}
Figure~\ref{fig:vgg16-lat-infer},~\ref{fig:resnet-lat-infer},~\ref{fig:xformer-lat-infer} demonstrate the inference latency improvement of MPC-Pipe for VGG16, ResNet, and Transformer. The left part of each figure shows the latency improvement in the LAN setting, while the right part shows the latency improvement in the WAN setting. In the LAN setting, MPC-Pipe provides a latency improvement of $10.9\%$, $10.6\%$, and $10.6\%$ for VGG16, ResNet, and Transformer, respectively. In the WAN setting, MPC-Pipe provides a latency improvement of $9.5\%$, $10.6\%$, and $7.4\%$ for VGG16, ResNet, and Transformer, respectively. Linear operations' latency mainly benefits from the inter-layer pipeline to hide communication of weight metadata, whereas Softmax, Maxpooling, and ReLU's latency benefits mainly from the inner-layer pipeline. \revision{For example, for VGG16, MPC parties are able to hide $0.23$s of communication for the linear layers and hide $1.24$s of computation for the non-linear layers resulting in overall $\pcent{10.9}$ latency improvement.} When transitioning from the LAN to the WAN setting, latency improvement for VGG16 and ResNet stays roughly the same, and that of the Transformer decreases slightly. When moving from the LAN setting to the WAN setting, non-linear layers see fewer performance benefits, while linear layers continue to see significant gains. This is because all the linear layers are computation-dominant, and the increased communication costs from network latency result in more performance improvements. For non-linear layers, the reverse is true. Depending on the model architecture, for VGG16 and ResNet, the latency improvement stays roughly the same because the increase in performance improvement for linear layers roughly balances out the non-linear layer behavior, resulting in similar latency improvement. For Transformers, improvements from Linear layers did not balance out the non-linear layer overheads as well.

\subsubsection{Training}
MPC-Pipe's training latency improvement is on par with inference latency improvement. As training is a more compute-intensive process, MPC-Pipe applies inter-linear and inner-layer pipelines more effectively to reduce the latency of the individual layer's runtime. \revision{For example, for VGG16, MPC parties will hide $2$s for linear functions using inter-linear pipelines and $1.49$s for the non-linear function using inner-layer pipelines. Combining both pipelines, MPC-Pipe can achieve overall $\pcent{11.6}$ throughput improvement for the training process.}

\subsection{MPC-Pipe with Different Connection Speed}

\begin{table}[h]
\centering
\caption{Throughput of the baseline MPC \& MPC-Pipe w/ different LAN speed in inputs/minute}
\begin{tabular}{|l|rr|rr|rr|}
\hline
\multicolumn{1}{|c|}{}          & \multicolumn{2}{c|}{VGG16}                            & \multicolumn{2}{c|}{ResNet}                            & \multicolumn{2}{c|}{Xformer}                           \\ \hline \hline
\multicolumn{1}{|c|}{Inference} & \multicolumn{1}{c|}{MPC}  & \multicolumn{1}{c|}{Pipe} & \multicolumn{1}{c|}{MPC}   & \multicolumn{1}{c|}{Pipe} & \multicolumn{1}{c|}{MPC}   & \multicolumn{1}{c|}{Pipe} \\ \hline
10Gbps                          & \multicolumn{1}{r|}{4.21} & \textbf{6.31}                      & \multicolumn{1}{r|}{9.00}  & \textbf{13.55}                     & \multicolumn{1}{r|}{10.52} & \textbf{13.93}                     \\ \hline
20Gbps                          & \multicolumn{1}{r|}{\textbf{6.42}} & 8.84                      & \multicolumn{1}{r|}{\textbf{13.77}} & 19.93                     & \multicolumn{1}{r|}{\textbf{14.86}} & 18.94                     \\ \hline
40Gbps                          & \multicolumn{1}{r|}{8.14} & 10.13                     & \multicolumn{1}{r|}{18.70} & 24.43                     & \multicolumn{1}{r|}{18.07} & 22.30                     \\ \hline \hline
\multicolumn{1}{|c|}{Training}  & \multicolumn{1}{c|}{MPC}  & \multicolumn{1}{c|}{Pipe} & \multicolumn{1}{c|}{MPC}   & \multicolumn{1}{c|}{Pipe} & \multicolumn{1}{c|}{MPC}   & \multicolumn{1}{c|}{Pipe} \\ \hline 
10Gbps                          & \multicolumn{1}{r|}{1.81} & \textbf{2.15}                      & \multicolumn{1}{r|}{5.10}  & \textbf{6.75}                      & \multicolumn{1}{r|}{5.62}  & \textbf{6.79}                      \\ \hline
20Gbps                          & \multicolumn{1}{r|}{\textbf{1.88}} & 2.29                      & \multicolumn{1}{r|}{\textbf{5.61}}  & 7.36                      & \multicolumn{1}{r|}{\textbf{7.16}}  & 8.18                      \\ \hline
40Gbps                          & \multicolumn{1}{r|}{2.46} & 2.70                      & \multicolumn{1}{r|}{8.82}  & 10.21                     & \multicolumn{1}{r|}{8.07}  & 8.93                      \\ \hline
\end{tabular}
\label{tab:throuVspeed}
\end{table}

Table~\ref{tab:throuVspeed} shows the throughput of baseline MPC (labeled MPC)  and MPC-Pipe (labeled Pipe) for different models using different interconnection speeds in the LAN setting. The throughput of MPC-Pipe using 10Gbps is similar to the baseline MPC at 20Gbps. For instance, the throughput of VGG16 inference using 20Gbps is 6.31 inputs/minute, and the throughput of VGG16 inference on MPC-Pipe using 10Gbps is 6.42 inputs/minute. A similar observation holds for the baseline MPC using 40Gbps LAN and MPC-Pipe using 20Gbps LAN. We observed similar trends for MPC-Pipe with WAN interconnection. 

\begin{table}[h]
\centering
\caption{Latency of the baseline MPC \& MPC-Pipe w/ different LAN speed in seconds}
\begin{tabular}{|c|rr|rr|rr|}
\hline
          & \multicolumn{2}{c|}{VGG16}                             & \multicolumn{2}{c|}{ResNet}                            & \multicolumn{2}{c|}{Xformer}                           \\ \hline \hline
Inference & \multicolumn{1}{c|}{MPC}   & \multicolumn{1}{c|}{Pipe} & \multicolumn{1}{c|}{MPC}   & \multicolumn{1}{c|}{Pipe} & \multicolumn{1}{c|}{MPC}   & \multicolumn{1}{c|}{Pipe} \\ \hline
10Gbps    & \multicolumn{1}{r|}{14.25} & 12.7                      & \multicolumn{1}{r|}{6.66}  & 5.96                      & \multicolumn{1}{r|}{5.78}  & 5.17                      \\ \hline
20Gbps    & \multicolumn{1}{r|}{9.6}   & 8.59                      & \multicolumn{1}{r|}{4.36}  & 3.92                      & \multicolumn{1}{r|}{4.03}  & 3.67                      \\ \hline
40Gbps    & \multicolumn{1}{r|}{7.36}  & 6.71                      & \multicolumn{1}{r|}{3.21}  & 2.86                      & \multicolumn{1}{r|}{3.32}  & 3.02                      \\ \hline \hline
Training  & \multicolumn{1}{c|}{MPC}   & \multicolumn{1}{c|}{Pipe} & \multicolumn{1}{c|}{MPC}   & \multicolumn{1}{c|}{Pipe} & \multicolumn{1}{c|}{MPC}   & \multicolumn{1}{c|}{Pipe} \\ \hline
10Gbps    & \multicolumn{1}{r|}{33.10} & 29.60                     & \multicolumn{1}{r|}{11.77} & 10.42                     & \multicolumn{1}{r|}{10.68} & 9.70                      \\ \hline
20Gbps    & \multicolumn{1}{r|}{31.88} & 29.34                     & \multicolumn{1}{r|}{10.70} & 9.68                      & \multicolumn{1}{r|}{8.39}  & 7.83                      \\ \hline
40Gbps    & \multicolumn{1}{r|}{24.38} & 23.02                     & \multicolumn{1}{r|}{6.80}  & 6.29                      & \multicolumn{1}{r|}{7.44}  & 7.05                      \\ \hline
\end{tabular}
\label{tab:latVspeed}
\end{table}

Table~\ref{tab:latVspeed} shows the latency of MPC and MPC-Pipe for different models using different interconnection speeds in the LAN setting. MPC-Pipe consistently outperforms the baseline. Particularly, MPC-Pipe with 20Gbps is only about 10\% slower than MPC baseline with 40Gbps.

\subsection{Scaling with more parties}
\textbf{Unscaled Bandwidth:} In this section, we will discuss the MPC-Pipe's performance benefits when we scale to more parties while keeping the bandwidth constant. Table~\ref{tab:3pclan} and~\ref{tab:3pcwan} describe MPC-Pipe's performance benefit in the 3PC setting with LAN and WAN connections. MPC-Pipe demonstrates noticeable throughput improvement by 52$\%$ and latency improvement by up to $16\%$ even when we did not scale the bandwidth availability with the number of parties.

\begin{table}[h]
    \centering
    \caption{MPC-Pipe's Performance Improvement 3PC with 10Gbps LAN.}
   \begin{tabular}{|ll|r|r|r|}
    \hline
    \multicolumn{2}{|l|}{}                                        & \multicolumn{1}{c|}{VGG16} & \multicolumn{1}{c|}{ResNet} & \multicolumn{1}{c|}{Xformer} \\ \hline \hline
    \multicolumn{1}{|l|}{\multirow{2}{*}{Inference}} & Throughput & $51\%$                     & $48\%$                      & $52\%$                       \\ \cline{2-5} 
    \multicolumn{1}{|l|}{}                           & Latency    & $13\%$                     & $10\%$                      & $14\%$                       \\ \hline\hline
    \multicolumn{1}{|l|}{\multirow{2}{*}{Training}}  & Throughput & $40\%$                     & $37\%$                      & $37\%$                       \\ \cline{2-5} 
    \multicolumn{1}{|l|}{}                           & Latency    & $16\%$                     & $16\%$                      & $13\%$                       \\ \hline
    \end{tabular}
        \label{tab:3pclan}
    \end{table}

\begin{table}[h]
    \centering
    \caption{MPC-Pipe's Performance Improvement 3PC with 10Gbps WAN.}
\begin{tabular}{|ll|r|r|r|}
\hline
\multicolumn{2}{|l|}{}                                        & \multicolumn{1}{c|}{VGG16} & \multicolumn{1}{c|}{ResNet} & \multicolumn{1}{c|}{Xformer} \\ \hline \hline
\multicolumn{1}{|l|}{\multirow{2}{*}{Inference}} & Throughput & $38\%$                     & $47\%$                      & $37\%$                       \\ \cline{2-5} 
\multicolumn{1}{|l|}{}                           & Latency    & $9.5\%$                    & $9.7\%$                     & $12\%$                       \\ \hline \hline
\multicolumn{1}{|l|}{\multirow{2}{*}{Training}}  & Throughput & $33\%$                     & $34\%$                      & $30\%$                       \\ \cline{2-5} 
\multicolumn{1}{|l|}{}                           & Latency    & $15\%$                     & $11\%$                      & $13\%$                       \\ \hline
\end{tabular}
    \label{tab:3pcwan}
\end{table}

\label{sec:more-parties}

\begin{figure}[h]
  \centering

    \includegraphics[width=7cm]{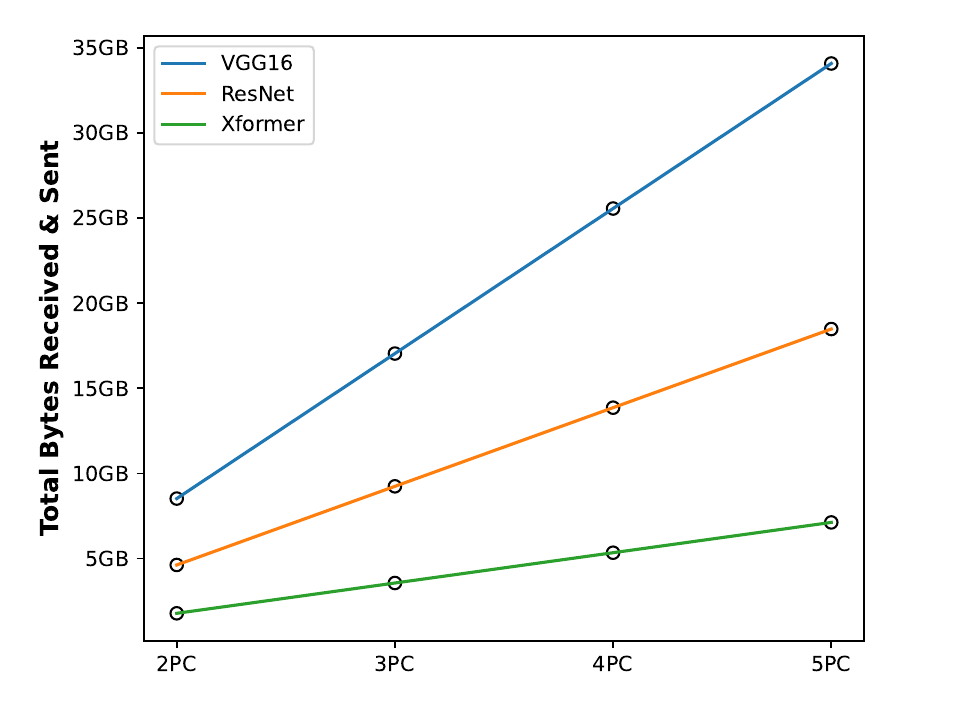}
  \caption{Total Bytes Sent and Received by an MPC Server.}
  \label{fig:bytes}
\end{figure}

\textbf{Scaled Bandwidth:} 
MPC-Pipe is built on generalized MPC protocols that are not limited to a specific number of parties. Hence, they can functionally scale to multiple MPC servers but their bandwidth demands grow. Figure~\ref{fig:bytes} plots the total bytes sent and received by an MPC server as we scale to more parties for inference. The number of total bytes increases linearly as we scale to more parties. In the 3PC setting for VGG16, roughly 17GB of data is sent and received by a single party, while in the 4PC setting, it is about 25GB. 
Thus, it is generally observed that as we scale to more parties, we will also have to scale the bandwidth between MPC servers. The question to answer is whether MPC-Pipe will continue to be beneficial when there is an increased bandwidth with the number of parties. To evaluate this question in this section for the 3PC setting, we assume the interconnection bandwidth is increased from 10Gbps to 20Gbps. Our results demonstrate that MPC-Pipe can still effectively overlap the computation and communication phases. The results in Table~\ref{tab:past} demonstrate this fact. The baseline MPC did improve its performance when using 3PC with 20Gbps, but MPC-Pipe was able to provide additional throughput and latency benefits. Compared to Table~\ref{tab:3pclan},  Resnet throughput throughput changed from 48$\%$ to 41$\%$ and latency changed from 10$\%$ improvement to 11$\%$ over baseline MPC. Thus MPC-Pipe continues to be useful even with increased bandwidth availability. 

Finally, we show the performance improvement of MPC-Pipe in the 2PC (with a 10Gbps interconnection) and 3PC (with a 20Gbps interconnection) settings in Table~\ref{tab:past}. MPC-Pipe's latency and throughput benefits remain roughly the same, scaling from 2PC to 3PC with scaled-up interconnection bandwidth. Based on these assumptions, we believe the MPC-Pipe's performance benefits with more MPC servers will remain stable.

\begin{table}[h]
\caption{MPC-Pipe Throughput (Tput) and Latency (Lat) Benefits for 2PC and 3PC w/ scaled up bandwidth over baseline MPC.}
\label{tab:past}
\begin{tabular}{|c|rr|rr|rr|}
\hline
           & \multicolumn{2}{c|}{VGG16}                             & \multicolumn{2}{c|}{ResNet}                            & \multicolumn{2}{c|}{Xformer}                           \\ \hline \hline
LAN        & \multicolumn{1}{c|}{Tput}   & \multicolumn{1}{c|}{Lat} & \multicolumn{1}{c|}{Tput}   & \multicolumn{1}{c|}{Lat} & \multicolumn{1}{c|}{Tput}   & \multicolumn{1}{c|}{Lat} \\ \hline
2PC 10Gb/s & \multicolumn{1}{r|}{49$\%$} & 11$\%$                   & \multicolumn{1}{r|}{50$\%$} & 11$\%$                   & \multicolumn{1}{r|}{34$\%$} & 11$\%$                   \\ \hline
3PC 20Gb/s & \multicolumn{1}{r|}{40$\%$} & 11$\%$                   & \multicolumn{1}{r|}{41$\%$} & 11$\%$                   & \multicolumn{1}{r|}{37$\%$} & 9.8$\%$                  \\ \hline \hline
WAN        & \multicolumn{1}{c|}{Tput}   & \multicolumn{1}{c|}{Lat} & \multicolumn{1}{c|}{Tput}   & \multicolumn{1}{c|}{Lat} & \multicolumn{1}{c|}{Tput}   & \multicolumn{1}{c|}{Lat} \\ \hline
2PC 10Gb/s & \multicolumn{1}{r|}{42$\%$} & 9.7$\%$                  & \multicolumn{1}{r|}{45$\%$} & 11$\%$                   & \multicolumn{1}{r|}{28$\%$} & 7.4$\%$                  \\ \hline
3PC 20Gb/s & \multicolumn{1}{r|}{38$\%$} & 9.4$\%$                  & \multicolumn{1}{r|}{41$\%$} & 12$\%$                   & \multicolumn{1}{r|}{29$\%$} & 8.1$\%$                  \\ \hline
\end{tabular}

\end{table}

\revision{\subsection{Offline phase}
MPC protocols adopt the offline/online computation paradigm, where MPC parties use the offline phase to prepare correlated random numbers, Beaver triples, which are not dependent on any inputs. MPC-Pipe achieves better throughput and latency for online phase execution without any additional costs to the offline phase. Our optimized online phase does not require new data that needs to be generated in the offline phase. Thus, MPC-Pipe's offline phase is identical to our baseline.}

\revision{Prior works have shown that offline phase can take up to $88\%$ of online$+$offline phase runtime~\cite{asplos23character}. Thus, a 50\% throughput improvement of the online phase may only improve overall throughput by 6\% if one were to consider the offline phase to be on the execution path. Because of MPC's offline/online computation paradigm, the offline phase, even if it is long, can be done beforehand.  Consequently, the offline phase latencies can be managed in such a way as not to impact the end-to-end latency since the offline phase is executed outside of the critical path. Many prior works and MPC-Pipe have thus focused on improving the online phase~\cite{reluMPC1, reluMPC2, reluMPC3, mpcformer}, which is a key for deployability.}

\subsection{Model accuracy}
\revision{Models evaluated using MPC-Pipe will have identical accuracy as if models are evaluated using the baselines. The inter-linear pipeline does not modify the linear layers' final computation results; it simply shifts the broadcasting to an earlier stage. The inner-layer pipeline breaks large vector operations into tiles, and all tiles will synchronize at the end of layer execution. Thus, execution using the inner-layer pipeline will provide identical results as if the inner-layer pipeline isn't used. For inter-batch pipelines, we break the original batch itself into 2 mini-batches and synchronize after each batch computes its gradients. Consequently, the gradient will be identical if inter-batch pipelines are not used. With all three pipeline schemes not changing the final computation results, the usage of MPC-Pipe will not impact the model's accuracy.}

%% file: charts/throughput_chart.tex
\begin{figure*}[htbp]
  \centering
  \begin{subfigure}[tb]{0.33\linewidth}
    \includegraphics[width=\linewidth]{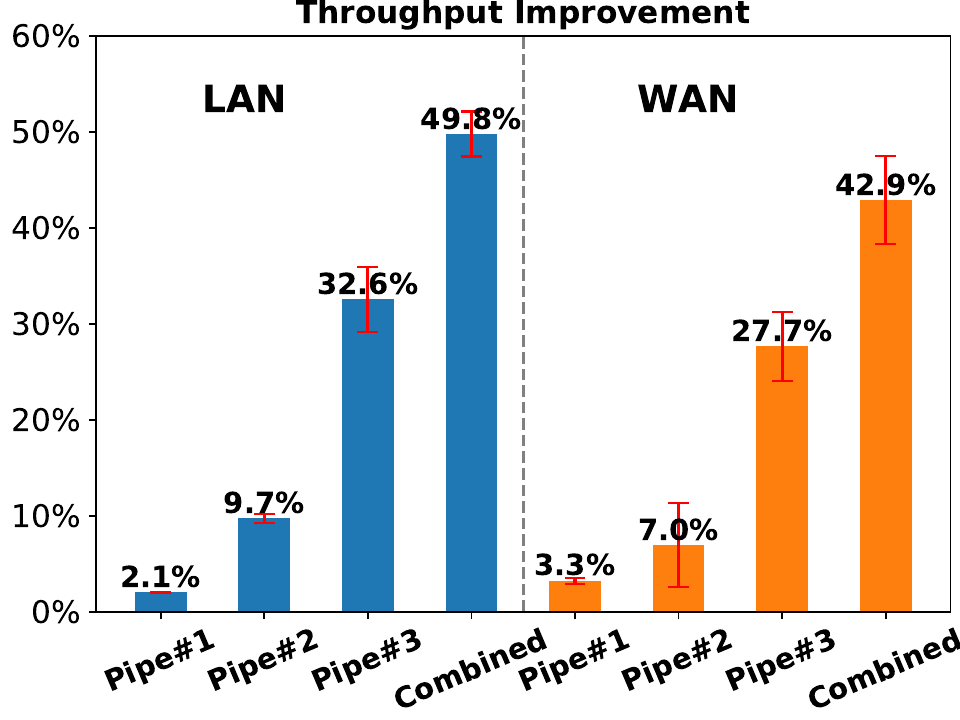}
    \caption{VGG16 Inference}
    \label{fig:vgg16-throu-lan}
  \end{subfigure}
  \begin{subfigure}[tb]{0.33\linewidth}
    \includegraphics[width=\linewidth]{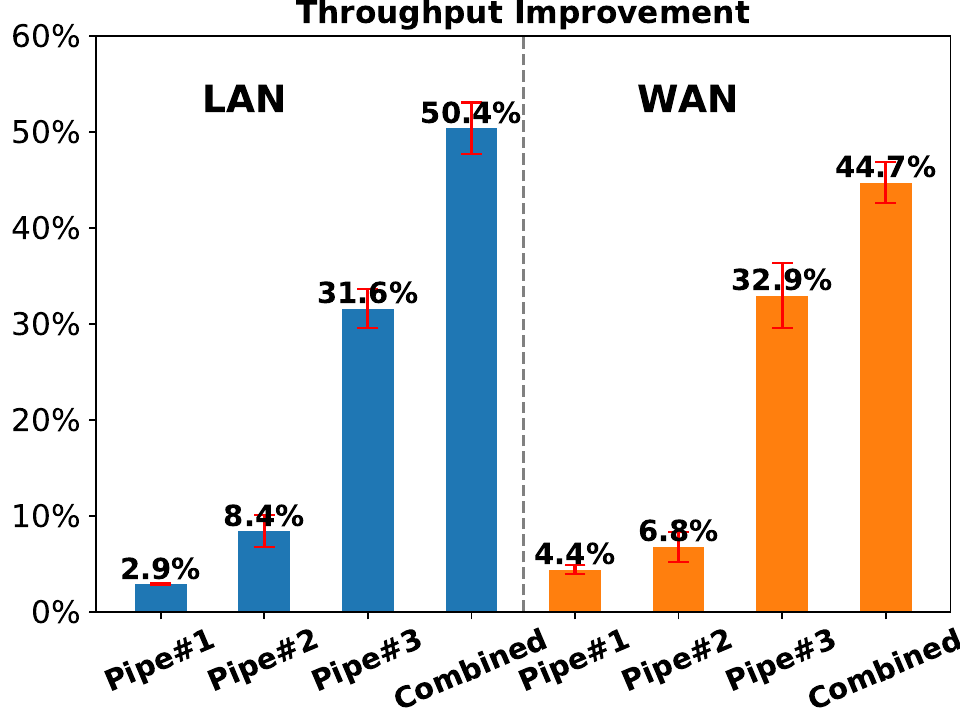}
    \caption{ResNet Inference}
    \label{fig:resnet-throu-lan}
  \end{subfigure}
    \begin{subfigure}[tb]{0.33\linewidth}
    \includegraphics[width=\linewidth]{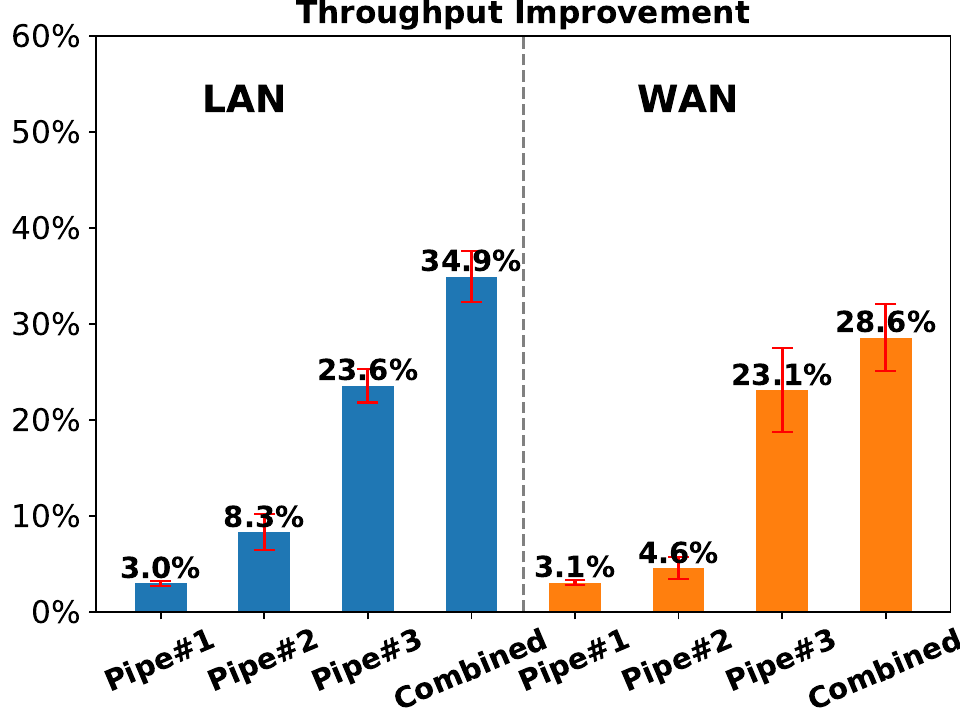}
    \caption{Transformer Inference}
    \label{fig:xformer-throu-lan}
  \end{subfigure}
  \label{fig:worst-speedups}
  \begin{subfigure}[tb]{0.33\linewidth}
    \includegraphics[width=\linewidth]{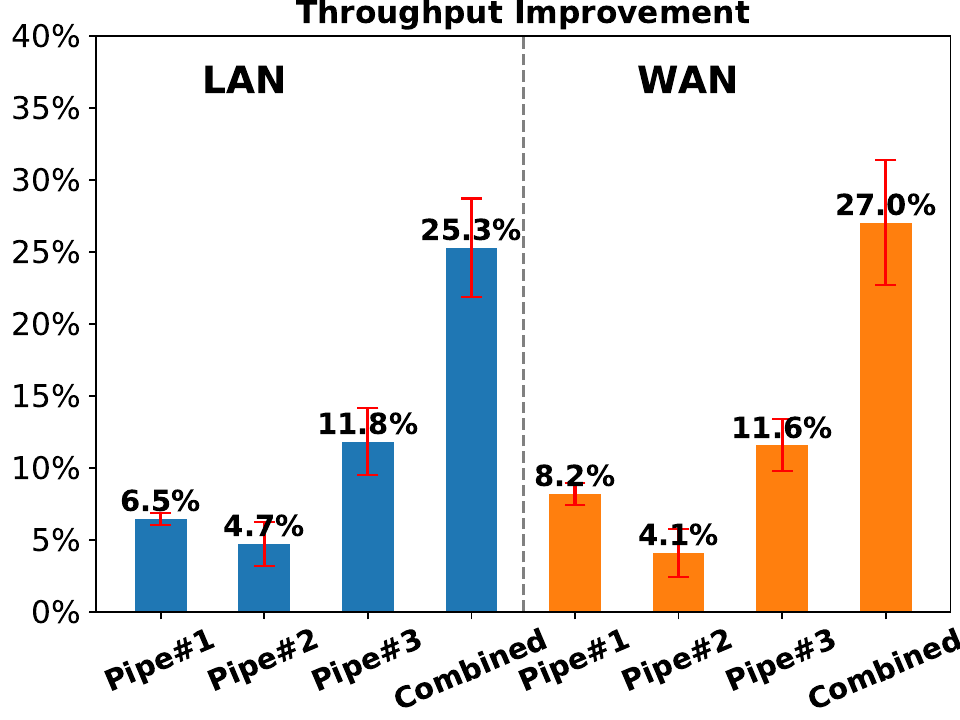}
    \caption{VGG16 Training}
    \label{fig:vgg16-throu-wan}
  \end{subfigure}
  \begin{subfigure}[tb]{0.33\linewidth}
    \includegraphics[width=\linewidth]{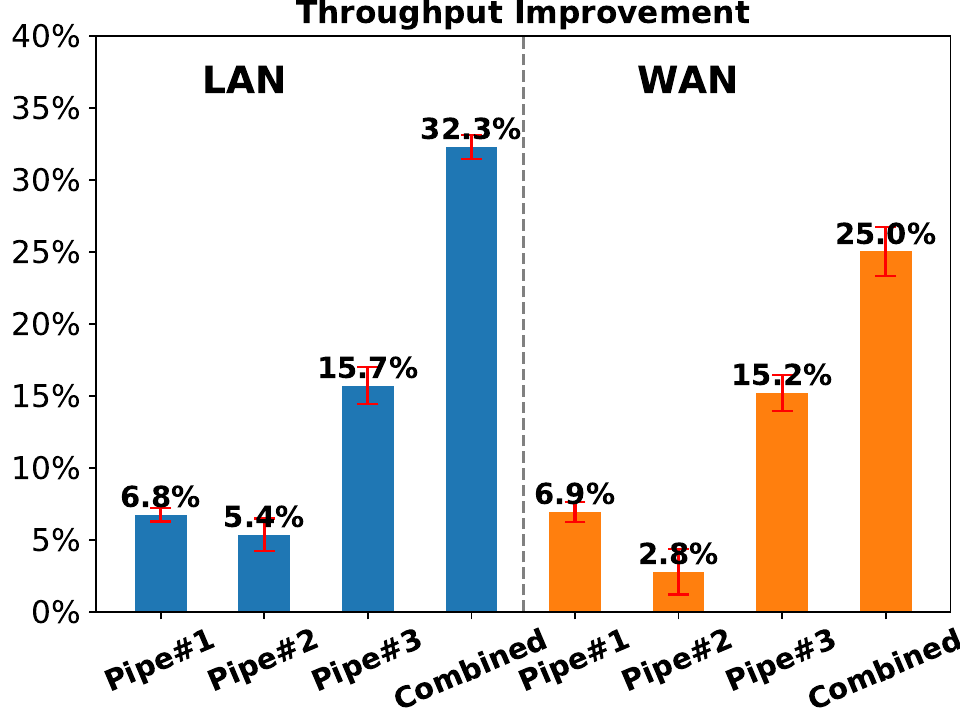}
    \caption{ResNet Training}
    \label{fig:resnet-throu-wan}
  \end{subfigure}
  \begin{subfigure}[tb]{0.33\linewidth}
    \includegraphics[width=\linewidth]{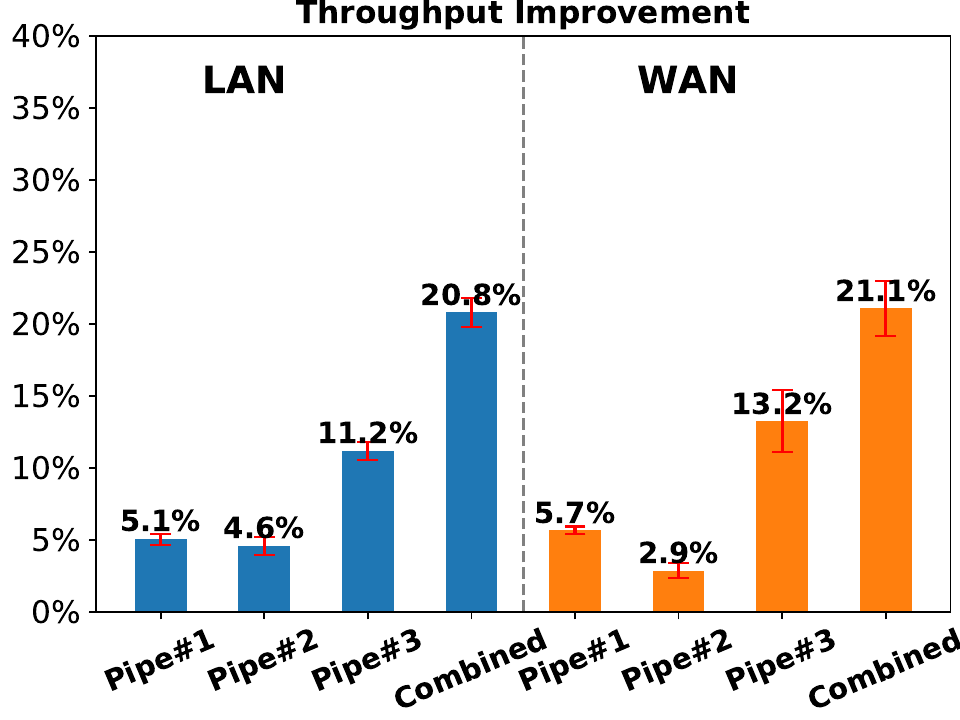}
    \caption{Transformer Training}
    \label{fig:xformer-throu-wan}
  \end{subfigure}
  \caption{MPC-Pipe 2PC Throughput Performance Benefits.}
  \label{fig:throuputs}
\end{figure*}

%% file: charts/throuput_categories.tex
\begin{table}[]
\caption{MPC-Pipe's throughput improvement for linear and non-linear layers of 2PC LAN.}
\label{tab:throu-cate}
\begin{tabular}{|l||rr|rr|}
\hline
\multirow{2}{*}{} & \multicolumn{2}{c||}{Inference}                                   & \multicolumn{2}{c|}{Training}                                    \\ \cline{2-5} 
                  & \multicolumn{1}{c|}{Linear}    & \multicolumn{1}{c||}{Non-Linear} & \multicolumn{1}{c|}{Linear}    & \multicolumn{1}{c|}{Non-Linear} \\ \hline
VGG16             & \multicolumn{1}{r|}{6.53$\%$}  & \multicolumn{1}{r||}{15.29$\%$}                     & \multicolumn{1}{r|}{9.89$\%$}  & 15.61$\%$                       \\ \hline
ResNet            & \multicolumn{1}{r|}{12.15$\%$} & \multicolumn{1}{r||}{11.69$\%$}                     & \multicolumn{1}{r|}{16.98$\%$} & 9.98$\%$                       \\ \hline
Xfomer            & \multicolumn{1}{r|}{8.42$\%$}  & \multicolumn{1}{r||}{14.02$\%$}                     & \multicolumn{1}{r|}{9.08$\%$}  & 11.85$\%$                       \\ \hline
\end{tabular}
\end{table}

%% file: charts/latency_chart.tex
\begin{figure*}[htbp]
  \centering
  \begin{subfigure}[tb]{0.33\linewidth}
    \includegraphics[width=\linewidth]{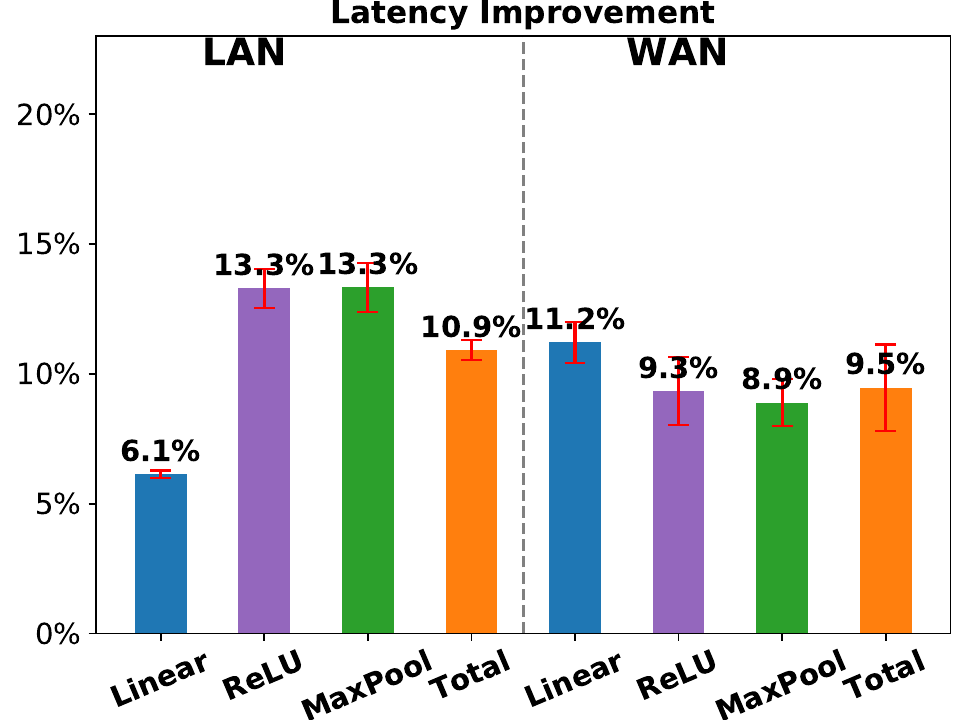}
    \caption{VGG16 Inference}
    \label{fig:vgg16-lat-infer}
  \end{subfigure}
  \begin{subfigure}[tb]{0.33\linewidth}
    \includegraphics[width=\linewidth]{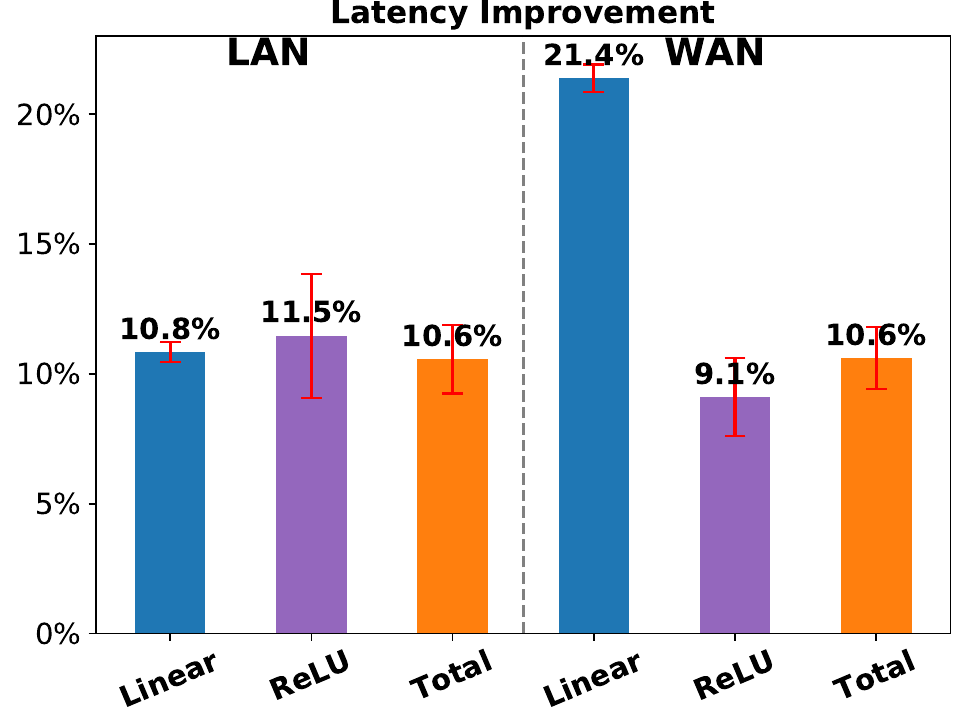}
    \caption{ResNet Inference}
    \label{fig:resnet-lat-infer}
  \end{subfigure}
    \begin{subfigure}[tb]{0.33\linewidth}
    \includegraphics[width=\linewidth]{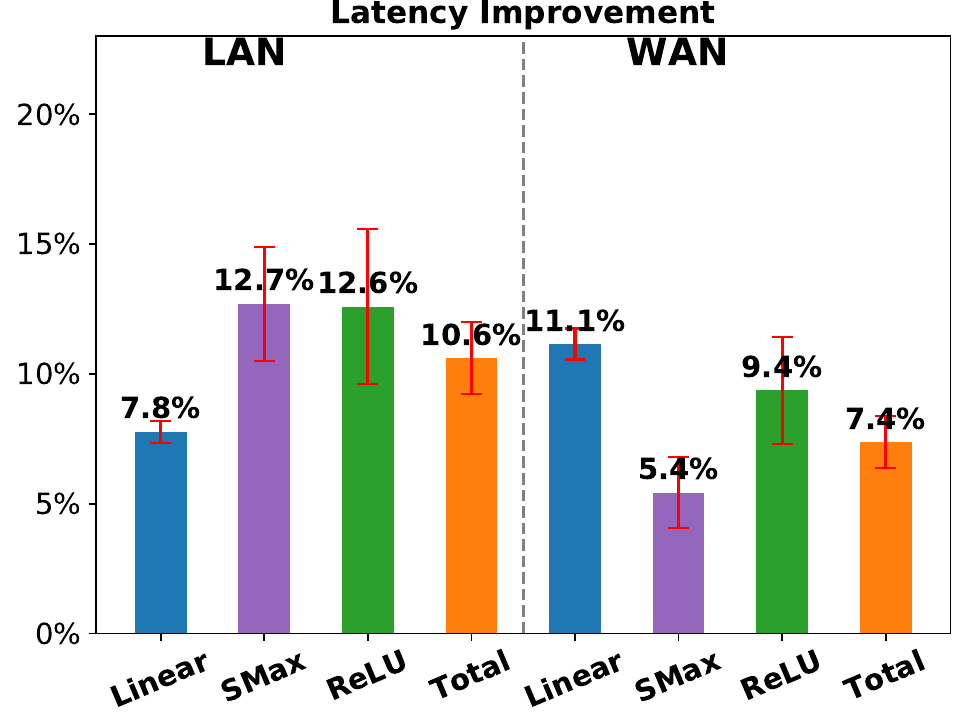}
    \caption{Transformer Inference}
    \label{fig:xformer-lat-infer}
  \end{subfigure}
  \label{fig:worst-speedups}
  \begin{subfigure}[tb]{0.33\linewidth}
    \includegraphics[width=\linewidth]{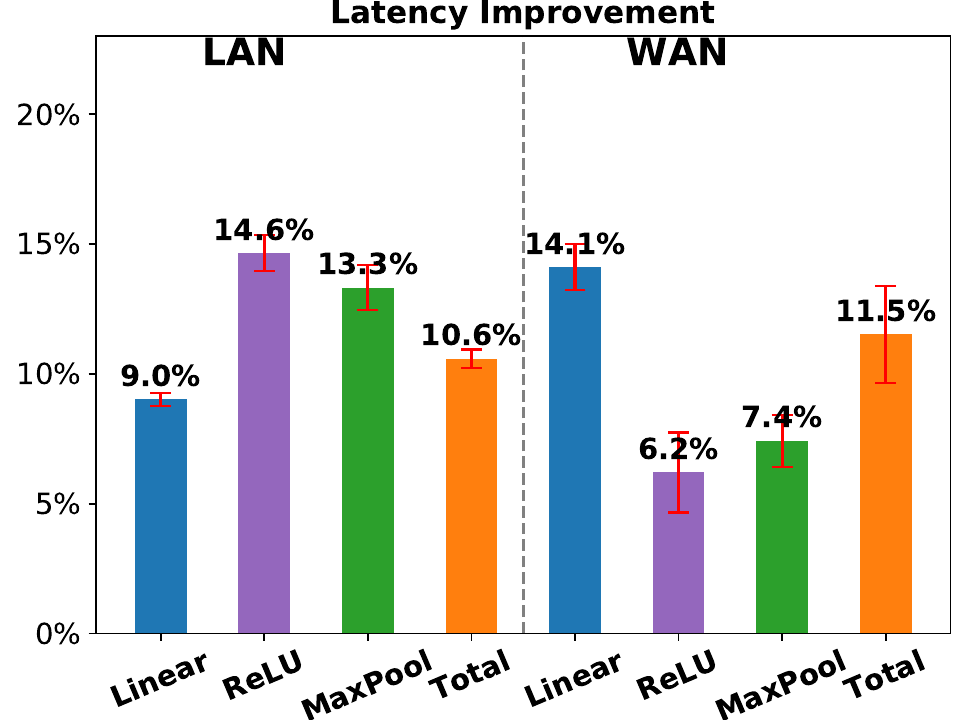}
    \caption{VGG16 Training}
    \label{fig:vgg16-lat-train}
  \end{subfigure}
  \begin{subfigure}[tb]{0.33\linewidth}
    \includegraphics[width=\linewidth]{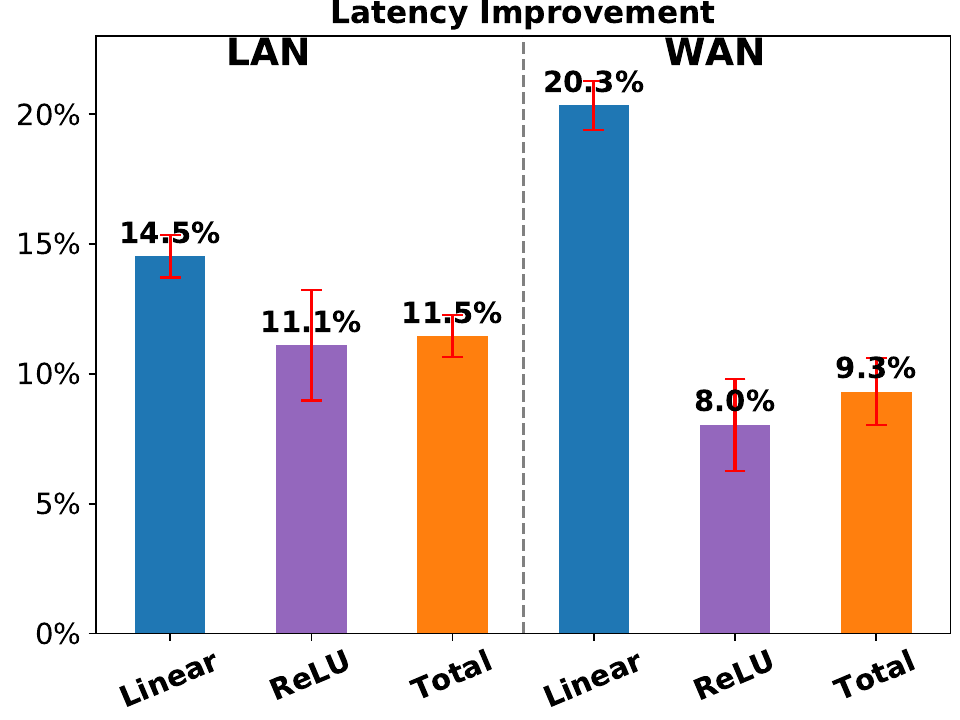}
    \caption{ResNet Training}
    \label{fig:resnet-lat-train}
  \end{subfigure}
  \begin{subfigure}[tb]{0.33\linewidth}
    \includegraphics[width=\linewidth]{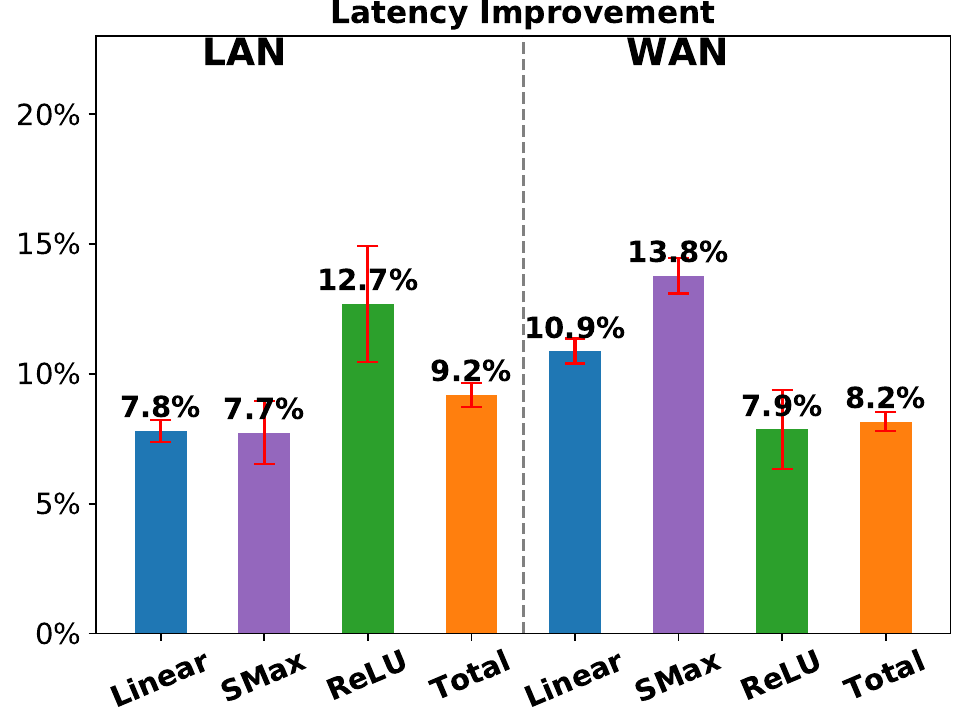}
    \caption{Transformer Training}
    \label{fig:xformer-lat-train}
  \end{subfigure}
  \caption{MPC-Pipe 2PC Latency Performance Benefits.}
  \label{fig:latency}
\end{figure*}

%% file: 6.conclusion.tex
In this work, we make a key observation that when MPC servers are transmitting MPC metadata for ML workloads, either GPU utility or network utilization is very low. We propose MPC-Pipe to overlap some of the computations with communications. MPC-Pipe is an efficient compute-communication pipeline scheme for secure MPC. MPC-Pipe proposes three key pipeline schemes, and our experiments reveal that MPC-Pipe significantly improves throughput and latency in MPC for both ML inference and training, highlighting its efficiency in secure ML applications.
